\documentclass[preprint, 3p, 11pt]{elsarticle}

\usepackage{graphicx}
\usepackage{amssymb}
 \usepackage{amsthm}
 \usepackage{amsmath}
 \usepackage{enumerate}
\usepackage{amssymb}
\usepackage{array}
\usepackage{booktabs}
\usepackage[ruled,vlined]{algorithm2e}
\SetAlCapSkip{1em}
\SetKwInput{KwInput}{Input}
\SetKwInput{KwOutput}{Output}
\usepackage{float}

\usepackage{scrextend}

\newsavebox{\mycode}
\newcommand{\vect}[1]{\text{$\mbox{\boldmath{$#1$}}$}}

\usepackage{epstopdf}

\journal{Structural Safety}

\begin{document}

\begin{frontmatter}



\title{Hamiltonian Monte Carlo Methods for Subset Simulation in Reliability Analysis}


\author[GUC]{Ziqi Wang}
\ead{ziqidwang@yahoo.com}

\author[IBK,SCCER]{Marco Broccardo}
\ead{bromarco@ethz.ch}

\author[SNU]{Junho Song\corref{cor1}}
\ead{junhosong@snu.ac.kr}
\cortext[cor1]{Corresponding author}

\address[GUC]{Earthquake Engineering Research and Test Center, Guangzhou University, China}
\address[IBK]{Institute of Structural Engineering, ETH Zurich, Switzerland}
\address[SCCER]{Swiss Competence Center for Energy Research---Supply of Electricity, ETH Zurich}
\address[SNU]{Department of Civil and Environmental Engineering, Seoul National University, South Korea}
\begin{abstract}
This paper studies a non-random-walk Markov Chain Monte Carlo method, namely the Hamiltonian Monte Carlo (HMC) method in the context of Subset Simulation used for structural reliability analysis. The HMC method relies on a deterministic mechanism inspired by Hamiltonian dynamics to propose samples following a target probability distribution. The method alleviates the random walk behavior to achieve a more effective and consistent exploration of the probability space compared to standard Gibbs or Metropolis-Hastings techniques. After a brief review of the basic concepts of the HMC method and its computational details, two algorithms are proposed to facilitate the application of the HMC method to Subset Simulation in structural reliability analysis. Next, the behavior of the two HMC algorithms is illustrated using simple probability distribution models. Finally, the accuracy and efficiency of Subset Simulation employing the two HMC algorithms are tested using various reliability examples. The supporting source code and data are available for download at (the URL that will become available once the paper is accepted).
\end{abstract}


\end{frontmatter}


\section{Introduction}
\label{sec: intro}
\noindent
Since analytical solutions of general reliability problems either at component or system level are usually unavailable, approximate reliability methods such as first- and second-order reliability methods \cite{ditlevsen1996str, der2005first}, response surface methods  \cite{faravelli1989resp, bucher1990fast}, and Monte Carlo simulation (MCS) techniques  \cite{rubinstein1998modern, rubinstein2013cross,  Au:2001aa, kurtz2013cross} have gained wide popularity. Compared with other reliability methods, MCS has the benefits of being accurate, insensitive to the complexity of limit-state functions and straightforward to implement. On the other hand, the efficiency of MCS depends on magnitude of the estimated probability. Since most practical reliability problems are characterized by small failure probabilities, the MCS scheme using the original probability density function can be computationally inefficient and often unfeasible. To enhance the efficiency of MCS, variance-reduction Monte Carlo methods have been developed. One powerful variance-reduction Monte Carlo method which has been widely used in reliability analysis is Subset Simulation \cite{Au:2001aa}. The method expresses the failure domain of interest as the intersection of a sequence of nested intermediate failure domains, and the failure probability of interest is expressed as a product of conditional probabilities associated with the intermediate failure domains. Since the conditional probabilities are significantly larger than the target failure probability the computational cost of Subset Simulation is significantly lower than the crude MCS method. The challenge of the scheme, which consists of evaluating the inter-mediate conditional probabilities, is overcome by using efficient Markov Chain Monte Carlo (MCMC) methods.  It is noted that approaches essentially similar to Subset Simulation have been independently developed for other statistical computing applications under the names, Sequential Monte Carlo method (or Particle Filters) \cite{cerou2012sequential} and Annealed Importance Sampling  \cite{neal2001annealed}. 	

	The crucial step in Subset Simulation is to obtain random samples according to a sequence of probability distributions that are conditional on nested intermediate failure domains. The efficien-cy and accuracy of Subset Simulation is directly affected by those of the MCMC algorithm used to produce random samples representing the conditional distributions in the sequence. In the current practice of Subset Simulation, various random-walk-based MCMC methods  \cite{miao2011modified, papaioannou2014mcmc} are employed to generate samples based on each conditional distribution model in the sequence. In this paper, a non-random-walk MCMC method, namely the Hamiltonian Monte Carlo (HMC) method \cite{duane1987hybrid, neal2011mcmc}, is studied in the context of Subset Simulation for reliability analysis. The HMC method employs a deterministic mechanism inspired by Hamiltonian dynamics to propose samples for a target probability distribution. The method alleviates the random-walk behavior to achieve a more effective and consistent exploration of the probability space compared to standard Gibbs or Metropolis-Hastings techniques.

	Originally developed in 1987 by Duane et al.  \cite{duane1987hybrid} under the name ``Hybrid Monte Carlo'' method for lattice field theory simulations in Lattice Quantum Chromodynamics, the HMC method has been introduced to mainstream statistical computing starting from the work by Neal  \cite{neal1993probabilistic} in 1993. The popularity of the HMC method has grown rapidly in recent years, and has proven a re-markable success in various statistical applications  \cite{neal2012bayesian, akhmatskaya2008gshmc, strathmann2015gradient}. However, to our knowledge, the application of the HMC to reliability analysis has never been studied. Motivated by this perspective, the paper studies the application of HMC in Subset Simulation for reliability analysis. In this context, the accuracy and efficiency of the HMC is investigated and compared with the conventional random-walk Metropolis-Hastings algorithm.

	The structure of this paper is as follows. Section \ref{sec: SS} briefly reviews the Subset Simulation. Section \ref{sec: HMC_p} introduces general concepts of HMC. Section \ref{sec:HMC4} develops the computational details of HMC algorithms for Subset Simulation method. Section \ref{sec:NI} shows the behavior of HMC-based Subset Simulation using simple distribution models. Next, in both standard normal space and non-Gaussian space, it is presented a series of numerical examples with analytical limit-state functions as well as structural reliability examples to test and demonstrate the validity of the method. Finally, Section \ref{sec:conc} presents a series concluding remarks and future directions.

\section{Principles of Subset Simulation}     \label{sec: SS}
\noindent
In reliability analysis, the failure probability of a system with basic random variables $x\in\mathbb{R}^n$ can be expressed by an integral,
\begin{equation}
 P_f  = \int_{\mathbb{R}^n}I_\mathcal{F} (\vect x) f(\vect x)d\vect x,\label{eq: p_f}
\end{equation}
where $I_\mathcal{F} (\cdot)$ is a binary indicator function which gives ``1'' if point $\vect x$ is within the failure domain, and ``0'' otherwise, and $f(\vect x)$ is the joint probability density function (PDF) of $\vect x$. A common practice in reliability analysis is to apply a transformation to random variables $\vect x$, denoted by $\vect x=T(\vect u)$, so that $\vect x$ can be expressed in terms of independent standard normal random variables $\vect u$. With the transformation, Eq.\eqref{eq: p_f} can be rewritten as
\begin{equation}
 P_f  = \int_{\mathbb{R}^n}I_\mathcal{F} [T(\vect u)]\varphi(\vect u)d\vect u,\label{eq: p_f_t}
\end{equation}
where $\varphi(\vect u)$ denotes the multivariate standard normal PDF. 

The Subset Simulation solution of Eq.(2) involves the construction of a sequence of nested intermediate failure domains, so that the failure domain of interest, $\mathcal{F}$, is expressed by
\begin{equation}
\mathcal{F} =\bigcap_{j=1}^M\mathcal{F}_j,\label{eq:3}
\end{equation}
where $\mathcal{F}_1\supset\mathcal{F}_2\supset \ldots\supset\mathcal{F}_M$ and $\mathcal{F} =\mathcal{F}_M$. The failure probability $P_f = \Pr(\vect u \in \mathcal{F})$ can be written as
\begin{equation}
\Pr(\vect u\in\mathcal{F}) = \prod_{j=1}^{M} \Pr(\vect u\in \mathcal{F}_j|\vect u \in \mathcal{F}_{j-1}),\label{eq:4}
\end{equation}
where $\Pr (\vect u \in F_0)=1$. Each $\Pr(\vect u\in F_{j-1})$ in Eq.\eqref{eq:4} can be computed using 
\begin{equation}
\Pr(\vect u\in \mathcal{F}_j|\vect u \in \mathcal{F}_{j-1}) = \int_{\mathbb{R}^n}I_{\mathcal{F}_j}(\vect u)\varphi(\vect u|\mathcal{F}_{j-1})d\vect u,\label{eq:5}
\end{equation}
where $\varphi(\vect u|\mathcal{F}_{j-1})$ is the conditional/truncated multivariate standard normal PDF. Using an MCMC technique to generate samples of $\varphi(\vect u|\mathcal{F}_{j-1})$, Eq.\eqref{eq:5} can be evaluated via MCS, i.e.
\begin{equation}
\Pr(\vect u\in \mathcal{F}_j|\vect u \in \mathcal{F}_{j-1}) \approx \frac{1}{N}\sum_{i=1}^NI_{\mathcal{F}_j}(\vect u_i),\label{eq:7}
\end{equation}
in which $\vect u_i$ are samples generated from conditional PDF $\varphi(\vect u|\mathcal{F}_{j-1})$ and $N$ the number of sample points.  In implementations of Subset Simulation, the nested failure domains are chosen adaptively such that $\Pr(\vect u \in \mathcal F_j |\vect u\in \mathcal{F}_{j-1})$, $j=1,2,\ldots M-1$, approximately equals to a specified percentile $p_0$. The estimator of the failure probability is then defined as follow 
\begin{equation}
 \hat P_f = \frac{p_0^{M-1}}{N}\sum_{i=1}^NI_{\mathcal{F}_M}(\vect u_i)\approx  P_f,\label{eq:7}
\end{equation}
where $\vect u_i$ are sampled from $\varphi(\vect u|\mathcal{F}_{M-1})$.

The estimator of the failure probability is biased because of the correlation of the samples \cite{Au:2001aa}, and the adaptive nature of the subsets \cite{cerou2012sequential}. The order of the bias $\mathcal{O}(N^{-1})$ is negligible compare to the coefficient of variation (c.o.v.), $\delta_f$, of the estimate. For a given run of the algorithm, an estimate of $\delta_{f}$ is given as \cite{Au:2001aa} 
\begin{equation}
	 \hat \delta_f \approx \sum_{i=1}^N\delta^2_j,\label{eq:8}
\end{equation}
where $\delta_j$ is the c.o.v of the $j^{th}$ subset which is given as follow
\begin{equation}
	 \delta_j  = \sqrt{\frac{1-P_j}{NP_j}(1+\gamma_j)},\label{eq:9}
\end{equation}
where $P_j=\Pr (\vect u \in \mathcal{F}_j |\vect u\in\mathcal{F}_{j-1})$ denotes the conditional probability, and $\gamma_j$ is expressed as
\begin{equation}
	\gamma_j  = 2\sum_{k=1}^{N/N_c-1}\left(1-\frac{kN_c}{N}\right)\rho_j(k),\label{eq:10}
\end{equation}
where $N_c=p_0N$ denotes the number of Markov chains at each subset level, and $\rho_j(k)$ is the average of the correlation coefficient at lag $k$ of the stationary sequence $\left[I_{\mathcal{F}_j}\left( \vect u_{j-1}^{\left( \frac{l-1}{p_o}+k\right)}\right), k=1,\ldots, N/N_c\right]$, $l = 1,\ldots,N_c,$ and $\rho_i(k)$ can be estimated directly from the sequence \cite{Au:2001aa}.
\section{General concepts of Hamiltonian Monte Carlo method}\label{sec: HMC_p}\noindent
This section provides a brief introduction of HMC method with the focus on its basic concepts, a detailed description of the method can be found in  \cite{duane1987hybrid, neal2011mcmc}. In specific, Section \ref{sec: HMC_p_1} introduces basic principles of Hamiltonian mechanics that are keys in formulating the HMC method. Then, Section \ref{sec: HMC_p_2} provides the ideas of HMC for sampling from a general distribution. 

\subsection{Hamiltonian mechanics} \label{sec: HMC_p_1}\noindent
Hamiltonian mechanics was proposed to provide a reformulation of classical mechanics in a more abstract form, but later it made significant contributions to the development of statistical mechanics and quantum mechanics. The Hamiltonian Monte Carlo method uses a deterministic procedure inspired by Hamiltonian mechanics to generate samples based on the target probability distribution. In this section a brief introduction of Hamiltonian mechanics is first provided. 
 
 Hamiltonian mechanics describes the time evolution of a system in terms of position vector $\vect q$ and momentum vector $\vect p$. The dimension of $\vect p$ and $\vect q$ should be identical, and $(\vect q, \vect p)$ defines a position-momentum phase space. The time evolution of the $(\vect q, \vect p)$ system is governed by \textit{Hamilton's equations} expressed by
 \begin{equation}
       \begin{split}
	\frac{d \vect q}{dt}&=\frac{\partial H}{\partial \vect p},\\
	\frac{d \vect p}{dt}&=-\frac{\partial H}{\partial \vect q},\\
	\frac{\partial H}{\partial t}&=-\frac{\partial L}{\partial t},
	\end{split}\label{eq:hamiltonian}
\end{equation} 
where $L=L(\vect q,\dot{\vect {q}},t)$ is the Lagrangian, which (in the non-relativistic setting) corresponds to the discrepancy between kinetic energy and potential energy, or free energy; and $H=H(\vect q,\vect p,t)$ is the Hamiltonian, which by definition is $H\equiv \vect p\cdot \vect q-L$. Eq.\eqref{eq:hamiltonian} works for both conservative and non-conservative systems, but HMC is formulated using Hamiltonian mechanics for conservative systems only, thus hereafter discussions on Hamiltonian mechanics are focused on conservative systems. 

For a system with only conservative forces the Lagrangian and Hamiltonian do not explicitly depend on time $t$, thus the last equation in Eq.\eqref{eq:hamiltonian}  can be dropped\footnote{For clarity, observe that the time dependence of $\vect q$ and $\vect p$ is implicitly assumed along the paper and it is made explicit only when it is necessary.}, leading to 
  \begin{equation}
       \begin{split}
	\frac{d \vect q}{dt}&=\frac{\partial H}{\partial \vect p},\\
	\frac{d \vect p}{dt}&=-\frac{\partial H}{\partial \vect q},
	\end{split}\label{eq:hamiltonian_c}
\end{equation} 
in which the Hamiltonian $H=H(\vect q,\vect p)$ is a constant corresponds to the total energy of the system, and thus $H$ is independent of time evolutions of (\vect q,\vect p). The Hamiltonian $H(\vect q,\vect p)$ can be expressed by
\begin{equation}
H(\vect q,\vect p) = V(\vect q) + K(\vect p) \label{H_d},
\end{equation}
where $V(\vect q)$ is the potential energy, which is a function of the position vector $\vect q$ alone, and $K(\vect p)$ is the kinetic energy, which is a function of the momentum vector $\vect p$ alone. Given initial values for the position and momentum, Eq.\eqref{H_d} completely defines the energy level for the system. Then, the solution of the Hamilton's equations, Eq.\eqref{eq:hamiltonian_c}, describes an equi-Hamiltonian trajectory of the system in the phase space. There are four fundamental properties of Hamiltonian dynamics (for conservative systems) that are keys to construct a valid MCMC method: 

\begin{enumerate}[\itshape i.]
\setlength{\itemindent}{-.1in}
\item \textit{Reversibility.} The mapping from a state $[\vect q(t),\vect p(t)]$ to a state $[\vect q(t+\tau),\vect p(t+\tau)]$ is an isomorphism (i.e. one-to-one); therefore, there always exists the unique inverse mapping. This is important in the context of MCMC because Hamiltonian dynamics can be used to construct reversible Markov chain transitions, which is a requirement to maintain the stationary distribution (hence the target distribution) invariant.  
\item \textit{Energy conservation.} The invariance of Hamiltonian has significant consequences in a MCMC method that uses a proposal arising from Hamiltonian dynamics. In this case, it is shown that the acceptance probability for the proposal is one, thus allowing an efficient and effective exploration of the probability space.
\item \textit{Volume conservation.} Volume preservation of Hamiltonian dynamics indicates if a mapping/ transformation governed by Hamilton's equations is applied to the points in some region of the phase space with volume $V$, after the transformation the image also has volume $V$. Volume preservation is implied from Eq.\eqref{eq:hamiltonian_c}, since it describes a shear transformation on the phase space, i.e. the determinant of the Jacobian of the transformation is one. This is important because a proposal arising from Hamiltonian dynamics does not need to account for the Jacobian of the transformation in the acceptance criterion of the MCMC.
\item \textit{Symplecticness.} Hamiltonian dynamics also conserves the sympletic structure of the phase space. A direct consequence of symplecticness is volume preservation. In the context of MCMC this has important implication on the choice of the numerical integrators used for solving Eq.\eqref{eq:hamiltonian_c}.

\end{enumerate}
\subsection{Hamiltonian Monte Carlo method} \label{sec: HMC_p_2}\noindent
To build a Monte Carlo method based on the deterministic Hamiltonian mechanics, one needs to establish a connection between the probability space of interest, and a mathematically equivalent Hamiltonian system described by time evolution of position and momentum. A probability space is defined by a sample space (defined by the set of all possible outcomes, $\vect x$), a set of events, and a PDF, $\pi(\vect x)$ (the paper restricts to continuous distributions with differentiable PDFs, so that HMC can be applied). 

To construct such a connection, first the outcome $\vect x$ is viewed as the position $\vect q$ of a Hamiltonian system (i.e. $\vect q\equiv \vect x$). Next, a set of auxiliary random momentum variables, $\vect p$, which has the same dimension as $\vect q$, are introduced to expand the original position space, so that now one has the position-momentum phase space of a Hamiltonian system. Finally, to incorporate the probabilistic structure of $\pi(\vect q)$ into the Hamiltonian system, the potential energy $V(\vect q)$ is defined in terms of the target PDF $\pi(\vect q)$ as	 
\begin{equation}
V(\vect q)\equiv -\log\pi(\vect q) \label{eq:V}.
\end{equation}
The form for kinetic energy $K(\vect p)$ could vary with implementation, but it is typically defined as 
\begin{equation}
 K(\vect p)  \equiv \frac{1}{2}(\vect p\cdot \mathcal{M}^{-1}\vect p) \label{eq:K},
\end{equation}
where $\mathcal{M}$ is a positive-definite and symmetric ``mass'' matrix. Typically $\mathcal{M}$ is chosen as a scalar multiple of the identity matrix. The joint PDF of $(\vect q,\vect p)$ is defined as
\begin{equation}
\pi(\vect q, \vect p) \equiv \frac{1}{Z}e^{-H(\vect q,\vect p)}= \frac{1}{Z}e^{-V(\vect q)}e^{-K(\vect p)}\label{eq:pdf},
\end{equation}
where $Z$ is a normalizing constant to make the density valid. In statistical mechanics, Eq.\eqref{eq:pdf} represents the canonical distribution \cite{neal2011mcmc}.

Substituting Eq.\eqref{eq:V} and Eq.\eqref{eq:K} into Eq.\eqref{eq:pdf}, one obtains
\begin{equation}
\pi(\vect q, \vect p) =\frac{1}{Z}e^{-H(\vect q,\vect p)}= \frac{1}{Z}\pi(\vect q)e^{-\frac{\vect p\cdot \mathcal{M}^{-1}\vect p}{2}}.\label{eq:pdf_2}
\end{equation}
The above definition of the joint PDF $\pi(\vect q, \vect p)$ has two important properties: a) the position and the momentum variables are statistically independent; and b) the position is distributed following the original target distribution $\pi(\vect q)$, and the momentum is distributed as a multivariate Gaussian distribution (given the kinetic energy defined by Eq.\eqref{eq:K}). The aforementioned two properties of $\pi(\vect q,\vect p)$ further suggests that if one could devise a method to sample from $\pi(\vect q,\vect p)$, then the method would readily obtain samples distributed as $\pi(\vect q)$ by simply projecting out the momentum component of $\pi(\vect q,\vect p)$ samples. In fact, HMC is a method to sample from $\pi(\vect q,\vect p)$. In particular, HMC sampling can be divided into two main steps. In the first step, the momentum is sampled from the canonical distribution; this together with the current position completely defines an equi-Hamiltonian hyper-surface. In the second step, both position and momentum variables change within the equi-Hamiltonian hyper-surface by integrating Eq. \eqref{eq:hamiltonian_c} for a given time $t_f$. The conceptual procedure of HMC is described in Algorithm \ref{alg:HMC}.\\
\begin{algorithm}[H]
\label{alg:HMC}
\begin{description}
\item[Step 1] Generate a random momentum $\vect p$ according to PDF $e^{-K(\vect p)}/Z$.
\item[Step 2] Use the momentum $\vect p$ and the position $\vect q$ of a seed sample as initial conditions, prop-\\ose a new state $(\vect q^*,\vect p^*)$ via solutions of the Hamilton's equations at a time point $t_f$.
\item[Step 3] Negate the proposed momentum, i.e., $\vect p^*\leftarrow-\vect p^*$.
\end{description}
 \caption{Conceptual procedure of Hamiltonian Monte Carlo method}
\end{algorithm}
$\ $\\
Note that Step 3 has no practical effects on Algorithm \ref{alg:HMC} and can be deleted in practice, since the momentum will be replaced by a random vector at the beginning of another run of the algorithm. However, Step 3 is conceptually important because with the negation the mechanism of proposing new states in HMC will be symmetric and therefore reversible. To see this, consider releasing the Hamiltonian system at $(\vect q,\vect p)$, after a specified duration the system deterministically reaches $(\vect q^*, \vect p^*)$. However, if the same system is released at $(\vect q^*,\vect p^*)$, after a same duration the system may not reach $(\vect q,\vect p)$, unless $\vect p^*$ is negated as suggested by Step 3.

The Step 1 in Algorithm 1 leaves the joint distribution of $(\vect q,\vect p)$ invariant, due to the independence of $\vect q$ and $\vect p$. The Step 2 combined with Step 3 in Algorithm 1 also leaves the joint distribution of $(\vect q,\vect p)$ invariant, due to the reversibility of the deterministic transition process and the invariance of Hamiltonian. To see this, consider the detailed balance equation
\begin{equation}
\pi[(\vect q^*,\vect p^*)]T[(\vect q,\vect p)|(\vect q^*,\vect p^*)]= \pi[(\vect q,\vect p)]T[(\vect q^*,\vect p^*)|(\vect q,\vect p)].\label{eq:det_bal}
\end{equation}

In the context of HMC, the transition between $(\vect q,\vect p)$  and $(\vect q^*,\vect p^*)$  in the phase space is a deterministic event with probability equals to either  0 or  1. Eq.\eqref{eq:det_bal} is satisfied if the transition probability is 0. On the other hand, if $(\vect q,\vect p)$  deterministically evolves to $(\vect q^*,\vect p^*)$ and vice versa (given the momentum negation), the detailed balance still holds due to the invariance of Hamiltonian (i.e. $\pi [\vect q^*,\vect p^*]= \pi [(\vect q,\vect p)]$).

If a numerical integration technique is used to solve Hamilton's equations, the invariance of the Hamiltonian can be violated, and consequently $\pi[(\vect q^*,\vect p^*)]\ne\pi[(\vect q,\vect p)]$. In that case a Metropolis accept-reject rule of the form $\min[1,\exp(-H(\vect q^*,\vect p^*)+H(\vect q,\vect p))]$ should be introduced to make Eq.\eqref{eq:det_bal} valid. Observe that the Metropolis accept-reject rule introduces a finite probability for the Hamiltonian system to remain at current state, which guarantees that the resulting chain is aperiodic. A mathematically more rigorous treatment for the detailed balance of HMC method can be found in \cite{neal2011mcmc}. Due to the deterministic nature of Hamiltonian dynamic systems, in general HMC alleviates random-walk behavior and explores the probability space in a consistent manner. 


\section{Hamiltonian Monte Carlo method for Subset Simulation}\label{sec:HMC4}
\noindent
The Algorithm \ref{alg:HMC} introduced in Section \ref{sec: HMC_p} provides a general framework of HMC. This section focuses on the implementation of HMC in the context of Subset Simulations, here denoted as HMC-SS. In specific, the first part of the section describes the HMC method to sample from truncated normal PDF $\varphi(\vect u|F_j)$ in the standard normal space, which is the classical setting for both the original Subset Simulation and structural reliability analysis, and the last part of the section focuses on the implementation of HMC to sample from truncated generic PDF $\pi(\vect q|F_j)$ in non-Gaussian spaces. 

\subsection{The Hamiltonian in the standard normal space}\label{sec:HMC_1}
\noindent
As discussed above, a common choice for the kinetic energy function in HMC is of the simple form $K(\vect p)=(\vect p \cdot \mathcal{M}^{-1}\vect p)/2$. It is observed in \cite{neal2011mcmc} that an ideal choice for $\mathcal{M}^{-1}$ is a matrix resembling the covariance of the target distribution. In this paper, $\mathcal{M}^{-1}$ is set to identity matrix (which is the covariance of $\varphi(\vect u)$). This kinetic energy selection indicates that the momentum $\vect p$ follows the multivariate standard normal distribution, denoted as $N(\vect 0,\vect I)$. Given this choice, the Hamiltonian $H(\vect u,\vect p)$ can be written as
 \begin{equation}
       \begin{split}
	H(\vect u,\vect p)&=V(\vect u)+K(\vect p)=-\log\varphi(\vect u|\mathcal{F}_j)+\frac{\vect p\cdot\vect p}{2},\\
	 			  &=\frac{\vect u\cdot\vect u}{2}+\frac{\vect p\cdot\vect p}{2}-\log I_{\mathcal{F}_j}(\vect u) + \text{const}.\\
	\end{split}\label{eq:hamiltonian_sn}
\end{equation} 

The constant term in Eq.\eqref{eq:hamiltonian_sn} can be dropped since it leaves Hamilton's equations intact. The term $-\log I_{\mathcal{F}_j}(\vect u)$ introduces a potential barrier to the system, so that proposals outside the failure domain $\mathcal{F}_j$ have infinite potential energy, that is, areas outside $\mathcal{F}_j$ cannot be reached by the Hamiltonian system.
\subsection{Solution of the Hamilton's equations}\label{sec:HMC_2}
\noindent 
As long as the trajectories of the Hamiltonian system lie in the failure domain $\mathcal{F}_j$, the term $-\log I_{\mathcal{F}_j}(\vect u)$ in Eq.\eqref{eq:hamiltonian_sn} is zero, and the Hamiltonian system has an analytical solution expressed by
 \begin{equation}
       \begin{split}
	\vect u(t)&=\vect p_{init}\sin t+\vect u_{init} \cos t,\\
	 \vect p(t)&=\vect p_{init}\cos t -\vect u_{init} \sin t,
	\end{split}\label{eq:hamilton_sol}
\end{equation} 
where $\vect p_{in}$ and $\vect u_{init}$  denote initial momentum and initial position, respectively.  In fact, it is easy to see that Eq.\eqref{eq:hamilton_sol} is also the analytical sampling formulation of HMC to sample from $\varphi(\vect u)$. Therefore, Eq.\eqref{eq:hamilton_sol} can be directly used to sample from $\varphi(\vect u|\mathcal{F}_j)$ if a rejection sampling technique is adopted, without considering how the Hamiltonian system interacts with the potential barrier.

 An alternative sampling approach, which accounts for the interaction with the potential barrier, is to introduce a bouncing mechanism, so that the system will bounce back to the failure domain when it hits the limit-state surface of $\mathcal{F}_j$. Algorithms and implementation details associated with these two approaches are     developed in the next section.

\subsection{Hamiltonian Monte Carlo algorithms for subset simulations}\label{sec:HMC_3}

\subsubsection{Rejection sampling based HMC (RS-HMC)}
\noindent
In terms of the difference in addressing the potential barrier in the Hamiltonian system, two algorithms of HMC method are proposed to sample from $\varphi(\vect u|\mathcal{F}_j)$. The first algorithm uses rejection sampling, and it is described in Algorithm \ref{alg:HMC_rej}.\\ 
\begin{algorithm}[H]
\label{alg:HMC_rej}
  \SetAlgoNoLine
\begin{description}
\item[Step 1] Generate a random initial momentum $\vect p_{init}$ according to $N(\vect 0,\mathcal{M})$. ($\mathcal{M}=I$ is used in this paper.)
\item[Step 2] Use the momentum $\vect p_{init}$ and the position $\vect u_{init}$ of a seed sample as initial conditions, propose a new state $(\vect u^*,\vect p^*)$ via Eq.\eqref{eq:hamilton_sol} at a specified time point $t_f$.
\item[Step 3] Acceptance criteria: \\
\eIf{$\vect u^*\in \mathcal{F}_j$ }{
	\begin{addmargin}[1em]{1em}
		accept the proposal.
	\end{addmargin}
        }{
        \begin{addmargin}[1em]{1em}
		set $\vect u^*$ to $\vect u_{init}$.
	\end{addmargin}}
\end{description}
 \caption{Rejection sampling based HMC to sample from $\varphi(\vect u | \mathcal{F}_j)$}
\end{algorithm}
$\ $\\
It is important to observe that the orbits described by Eq.\eqref{eq:hamilton_sol}  are ellipses, therefore periodic. This can, theoretically, lead to a periodic chain, which violates the ergodic property of the Markov transition. Note that ergodicity is a fundamental requirement to guarantee that the stationary distribution of the Markov chain is effectively the target distribution. However, the randomization of momentum variables, and the avoidance of setting $t_f=2\pi$ (note that  Eq.\eqref{eq:hamilton_sol} has a period of $2\pi$) would destroy the possible periodicity. 

Observe that step 3 can be reviewed as a MH acceptance criteria $\min[1,I_{\mathcal{F}_j}(\vect q^*)]$. Since the $I_{\mathcal{F}_j}(\vect q^*)$ is either 1 or 0 the acceptance rate has the same statistical properties of the indicator function $I_{\mathcal{F}_j}$. Note that these properties, however, are conditional to the initial samples belonging to the subset $j$. Moreover, in this context, the negation of the momentum has an elegant interpretation, that is: if the sample fails outside the failure domain, the momentum is inverted and the position returns back to its initial state. 

Notably, the HMC based Algorithm 2 is analogous to the algorithm proposed in \cite{papaioannou2014mcmc} with an isotropic cross-correlation matrix between the current state and the proposal. In the HMC context the role of cross-correlation matrix is played by the mass matrix $\mathcal{M}$. Observe that generally $\mathcal{M}$ can be non-isotropic; however, it must be symmetric and positive define. This corroborates the observation in \cite{au2016mcmc}, where it is shown that only symmetric cross-correlation matrices are valid for reversible MCMCs.

 In HMC-SS, the fundamental tuning parameter is the time $t_f$. In terms of Eq.\eqref{eq:hamilton_sol}, for a time point moves from $t=0$, to $t=\pi/2$,  $t=\pi$, $t=3\pi/2$, and finally to $t=2\pi$, the corresponding position vector moves from $\vect u_{init}$, to $\vect p_{init}$, $\vect u_{init}$, -$\vect p_{init}$, and finally return to $\vect u_{init}$. Considering this circulating behavior of Eq.\eqref{eq:hamilton_sol} in conjunction with the fact that the failure events are likely to be observed in the vicinity of $\vect u_{init}$, a reasonable choice for $t_f$ in Algorithm 2 is $t_f\in[-\pi/2, \pi/2]$. Moreover, with $\mathcal{F}_j$ becoming smaller, to have a relatively constant acceptance rate $t_f$ should be decreased accordingly. Motivated by this idea, an adaptive approach similar to \cite{papaioannou2014mcmc, haario2001adaptive} could be used to select $t_f$ in HMC-SS. The procedure of the adaptive approach to select $t_f$ is described in Algorithm \ref{alg:adaptrule}.\\
\begin{algorithm}[H]
  \SetAlgoNoLine
 \begin{description}
\item[Step 1]  For the current intermediate $j$ of Subset Simulation\;
  \begin{addmargin}[1em]{2em}
    	\eIf{$j=1$}{\begin{addmargin}[1em]{1em}
     		 initialize $t_f $ as, e.g., $t_f=\pi/4$\;
		  \end{addmargin}
    		}{\begin{addmargin}[1em]{2em}
    		  initialize $t_f$ as the $t_f$ selected in step $j-1$;\
   		 \end{addmargin}}
    \end{addmargin}
\item[Step 2]   Compute the acceptance rate, denoted as $a$, for every $N_a$ chains simulated. If $a<a_{low}$ (it is suggested $a_{low}=0.3)$, set $t_f\leftarrow\sin^{-1}\{\sin(t_f)\exp[(a-a_{low})/2]\}$. Similarly, if $a>a_{up}$ (it is suggested $a_{up}=0.5$), set $t_f\leftarrow\sin^{-1}\{\sin(t_f)\exp[(a-a_{up})/2]\}$. 
\end{description}
  \caption{Adaptive rule to select $t_f$}
  \label{alg:adaptrule}
\end{algorithm}
$\ $\\
Note that in each intermediate step of Subset Simulation, $p_0 N$ Markov chains are simulated (each chain has $1/p_0$  samples), one can divide $p_0N$ into an integer number of portions with each portion contains $N_a$ chains, and compute the acceptance rate for every $N_a$ chains. Also, note that the rule to adapt $t_f$ in Step 2 of Algorithm \ref{alg:adaptrule} is similar to the methods discussed in \cite{papaioannou2014mcmc, haario2001adaptive}. The adaptations of $t_f$ may not be the optimal one, but at least it provides the right trend to modify $t_f$: if the acceptance rate is too low, $t_f$ is decreased; if the acceptance rate is too high, $t_f$ is increased. 

By using Algorithm 3, one could have an HMC sequence with acceptance rate approximately ranged in $[a_{low}, a_{up}]$. However, an issue not addressed by Algorithm \ref{alg:adaptrule} is that for low probability levels, $t_f$ has to be fairly small to have a reasonable acceptance rate, leading to an increase of random walk behavior of HMC proposals. To address this issue, it is noted that the only cause of the random walk behavior in HMC algorithm is the randomization of momentum vector $\vect p_{init}$ at the beginning of each run of the algorithm, thus a \textit{partial momentum refreshment technique} \cite{horowitz1991generalized} can be introduced to suppress the random walk behavior. 
	
The partial momentum refreshment technique simply replaces the original random initial momentum $\vect p_{init}$ in Step 1 of Algorithm \ref{alg:HMC_rej} by a modified momentum, denoted by $\vect p_{init}'$, obtained using the following equation \cite{horowitz1991generalized}
 \begin{equation}
        \vect p'_{init} = \alpha \vect p^*+\sqrt{1-\alpha^2}\vect p_{init},
        \label{eq:pr}
\end{equation} 
where $\alpha \in [-1, 1]$, and $\vect p^*$ is the momentum at the end of previous HMC trajectory. An $\alpha$ value of 0 is associated with the case of starting a new trajectory by total randomization of the momentum, while an $\alpha$ value of $\pm1$ is associated with case of tracing/retracing the previous trajectory. How to tune parameter $\alpha$ in the context of Subset Simulation is an open question requiring further investigations, and is beyond the scope of this study. In this paper, a simple example is used to illustrate the influence of $\alpha$ on HMC sampling. 

\subsubsection{Barrier bouncing based HMC (BB-HMC)}
\noindent An alternative approach to address the potential barrier is to introduce a bouncing mechanism \cite{pakman2014exact} of the system. The shift in momentum during the bouncing is expressed by \cite{pakman2014exact}
\begin{equation}
\vect p_a = \vect p_b -2(\vect p_b\cdot\vect v)\vect v,
\end{equation}
where $\vect p_b$ denotes the momentum instantaneously before the system hits the potential barrier, $\vect p_a$ denotes the momentum instantaneously after the system hits the potential barrier, and the direction vector $\vect v$ is expressed by
\begin{equation}
\vect v = -\frac{\nabla_{\vect u} G}{||\nabla_{\vect u}G||},
\end{equation}
where $\nabla_{\vect u} G$ denotes the gradient of the potential barrier (or limit-state surface for this paper) at the bouncing point. With the bouncing mechanism, for a specified initial $\vect p_{init}$ and $\vect u_{init}$, and the time point $t_f$ at which new state $(\vect u^*,\vect p^* )$ is computed, one may need to determine the time point $t_h$, $t_h<t_f$, at which the system hits the barrier for the first time. Time point $t_h$ can be numerically determined using a secant or Newton-Raphson method described in Algorithm \ref{alg:sec} and \ref{alg:NR}.
\begin{algorithm}[!b]
  \SetAlgoNoLine
 \begin{description}
   \item[Step 1]  For the current state $(\vect u(t_0), \vect p(t_0))$, $t_0=0$, and a proposed state $(\vect u(t_i),\vect p(t_1))$, $t_1=t_f$, with  $G_{\mathcal{F}_j} (\vect u(t_1))> $\upshape{toll}, for $i = 1, 2,...$\ :
  \begin{addmargin}[1em]{2em}
  	\While{$|G_{\mathcal{F}_j} (\vect u(t_i))|>$\upshape{toll}} {
	\begin{addmargin}[1em]{2em}
		$\lambda\leftarrow 1$, evaluate:
		\begin{equation}
			t_i\leftarrow t_{i-1}-\lambda \frac{G_{\mathcal{F}_j}(\vect u(t_{i-1}))(t_{i-1}-t_{i-2})}{G_{\mathcal{F}_j}(\vect u(t_{i-1}))-G_{\mathcal{F}_j}(\vect u(t_{i-2}))}\nonumber
		\end{equation}
	
    	\While{$t_i<0 \wedge t_i>t_1$}{\begin{addmargin}[1em]{1em}
     		  $\lambda\leftarrow 0.5\lambda$, evaluate:\
		  \begin{equation}
			t_i\leftarrow t_{i-1}-\text{sign}(t_i)\lambda \frac{G_{\mathcal{F}_j}(\vect u(t_{i-1}))(t_{i-1}-t_{i-2})}{G_{\mathcal{F}_j}(\vect u(t_{i-1}))-G_{\mathcal{F}_j}(\vect u(t_{i-2}))}\nonumber
		\end{equation}
		  \end{addmargin}
    		}\end{addmargin}
	 }
    \end{addmargin}
  \item[Step 2]  Set $t_h\leftarrow t_i$. 
  \end{description}
  \caption{Computing $t_h$ using the secant method}
  \label{alg:sec}
\end{algorithm}
\begin{algorithm}[t]
  \SetAlgoNoLine
 \begin{description}
 \item[Step 1]  For the current state $(\vect u(t_0), \vect p(t_0))$, $t_0=0$, and a proposed state $(\vect u(t_i),\vect p(t_1))$, $t_1=t_f$, with  $G_{\mathcal{F}_j} (\vect u(t_1))>0$\ :
 \begin{equation}
t_2\leftarrow  \frac{t_0G_{\mathcal{F}_j}(\vect u(t_{1}))-t_1G_{\mathcal{F}_j}(\vect u(t_{0}))}{ G_{\mathcal{F}_j}(\vect u(t_{1}))-G_{\mathcal{F}_j}(\vect u(t_{0}))}\nonumber
 \end{equation}

   \item[Step 2]  For $i = 3, 4,...$\ :
  \begin{addmargin}[1em]{2em}
  	\While{$|G_{\mathcal{F}_j} (\vect u(t_i))|>$ \upshape{toll}} {
	\begin{addmargin}[1em]{2em}
		$\lambda\leftarrow 1$, evaluate:
		\begin{equation}
			t_i\leftarrow t_{i-1}-\lambda \frac{G_{\mathcal{F}_j}(\vect u(t_{i-1}))}{G'_{\mathcal{F}_j}(\vect u(t_{i-1}))}=t_{i-1}-\lambda \frac{G_{\mathcal{F}_j}(\vect u(t_{i-1}))}{\nabla_{\vect u}G_{\mathcal{F}_j}(\vect u(t_{i-1}))\cdot \mathcal{M}^{-1}\vect p(t_{i-1})}\nonumber
		\end{equation}
	
    	\While{$t_i<0 \wedge t_i>t_1$}{\begin{addmargin}[1em]{1em}
     		  $\lambda\leftarrow 0.5\lambda$, evaluate:\
		  \begin{equation}
			t_i\leftarrow t_{i-1}-\text{sign}(t_i)\lambda \frac{G_{\mathcal{F}_j}(\vect u(t_{i-1}))}{\nabla_{\vect u}G_{\mathcal{F}_j}(\vect u(t_{i-1}))\cdot \mathcal{M}^{-1}\vect p(t_{i-1})}\nonumber
		\end{equation}
		  \end{addmargin}
    		}\end{addmargin}
	 }
    \end{addmargin}
  \item[Step 3]  Set $t_h\leftarrow t_i$. 
  \end{description}
  \caption{Computing $t_h$ using Newton-Raphson method}
  \label{alg:NR}
\end{algorithm}

Note that the secant update step in Step 1 of Algorithm 5 is introduced to accelerate the Newton-Raphson method, and it does not introduce additional limit-state function evaluations since initially one has to propose a state $(\vect u(t_1), \vect p(t_1))$ to determine if a bouncing process needs to be simulated. Also, note that due to the chain rule, the time derivative of $G(\vect u(t_i ))$ is written as 
\begin{equation}
\frac{dG(\vect u(t))}{dt} = \nabla_{\vect u} G(\vect u(t))\cdot \mathcal{M}^{-1}\vect p(t),
\end{equation}
where it is used $d\vect u(t)/dt=\mathcal{M}^{-1} \vect p(t)$. This property becomes particularly convenient when the Hamilton's equations are solved numerically. Notice that when the gradient $\nabla_{\vect u}G(\vect u(t))$ is not directly available, it can be computed with an efficient scheme like the direct differentiation method (DDM) \cite{zhang1993dynamic}.

In the context of this paper,  $t_f$ is typically within $[-\pi/2, \pi/2]$ (i.e., a quarter of the Hamiltonian system period); moreover, the step size $\lambda$  introduced in Algorithm 4 and Algorithm 5 corrects the trail points of $t_h$ if they are smaller than 0 or larger than $t_f$. It follows that the secant or Newton-Raphson method find the time point  that corresponds to the first hit of the barrier. With $t_h$ obtained from the secant or Newton-Raphson method, the state $(\vect u_b, \vect p_b )$ instantaneously before the bouncing is obtained as
 \begin{equation}
       \begin{split}
	\vect u_b&=\vect p_{init}\sin t_h+\vect u_{init} \cos t_h,\\
	 \vect p_b&=\vect p_{init}\cos t_h -\vect u_{init} \sin t_h.
	\end{split}\label{eq:hamilton_sol_1}
\end{equation} 
The state $(\vect u_a,\vect p_a )$ instantaneously after the bouncing is obtained as 
\begin{equation}
       \begin{split}
	\vect u_a&=\vect u_b,\\
	 \vect p_a&=\vect p_{b}-2(\vect p_b\cdot\vect v)\vect v.
	\end{split}\label{eq:hamilton_sol_2}
\end{equation} 
Note that the direction vector $\vect v$ requires the gradient of the limit-state surface. If a gradient-free BB-HMC method is required, one could use the secant method to solve for $t_h$ and replace vector $\vect v$ in \eqref{eq:hamilton_sol_2} by a vector $\hat{\vect v}$ obtained as
\begin{equation}
\hat{\vect v} = \frac{\vect u(t_{i^*})-\vect u(t_{i^*-1})}{||\vect u(t_{i^*}-\vect u(t_{i^*-1})||},
\label{eq:v_approx}
\end{equation} 
in which $\vect u(t_{i^* } )$ and $\vect u(t_{i^*-1})$ correspond to the position vectors of the last two iterations of Algorithm \ref{alg:sec}. Eq.\eqref{eq:v_approx} is proposed based on the assumption that the secant direction may crudely approximate the tangent direction in Newton's method. After Eq.\eqref{eq:v_approx}, the direction of $\hat{\vect v}$ may require a correction expressed by
\begin{equation}
\begin{split}
&\mathbf{if\  } \vect p_p\cdot\hat{\vect v}>0, \mathbf{\ then\  } \\
 &\ \ \ \ \hat{\vect v} \leftarrow -\hat{\vect v} \\
&\mathbf{end}.  
\end{split}
\label{eq:v_approx2}
\end{equation} 
Clearly, if the Newton-Raphson method is used to solve for $t_h$ (which implies one could obtain the gradient of the limit-state function), the use of $\hat{\vect v}$ can be avoided. 

Finally, after the bouncing, the system proceeds as
 \begin{equation}
       \begin{split}
	\vect u_b&=\vect p_{a}\sin (t-t_h)+\vect u_{a} \cos (t-t_h),\\
	 \vect p_b&=\vect p_{a}\cos (t-t_h) -\vect u_{a} \sin (t-t_h),
	\end{split}\label{eq:hamilton_sol_bounc}
\end{equation} 
until it hits the potential barrier again, and the same aforementioned procedure can be applied again. In practical implementation of BB-HMC, due to the additional computational cost introduced in computing $t_h$, for each proposal one should avoid the computation of $t_h$ for a second and more time. This can be accomplished by: i. using an adaptive rule similar to Algorithm \ref{alg:adaptrule}, starting with $t_f=\pi/4$, and compute the acceptance rate, $a$, for each $N_a$ chains simulated, if $a<a^* $(it is suggested $a^*=0.8$), set $t_f\leftarrow\sin^{-1}\{\sin(t_f)\exp[(a-a_{low})/2]\}$; and ii. reject the proposal if it is not in the failure domain, without simulating the bouncing process for a second time. 

Using the ideas introduced in this section, BB-HMC algorithm is developed as in Algorithm \ref{alg:HMCBB}. 
\begin{algorithm}[!b]
  \SetAlgoNoLine
\label{alg:HMCBB}
\begin{description}
\item[Step 1] Generate a random initial momentum $\vect p_{init}$ according to $N(\vect 0,\mathcal{M})$, ($\mathcal{M}=I$ is used in this paper). 
\item[Step 2] . Using momentum $\vect p_{init}$ and position $\vect u_{init}$ of a seed sample as initial conditions,  \\ propose a new state $(\vect u^*,\vect p^* )$ via Eq.\eqref{eq:hamilton_sol} with an adaptively selected time point $t_f. $ 
\item[Step 3] Accept the proposal if $\vect u^*\in\mathcal{F}_j$, otherwise go to Step 4.
\item[Step 4] Use $(\vect u^*,\vect p^* )$ and $(\vect u_{init},\vect p_{init} )$ and their corresponding limit-state function values as initials of Algorithm \ref{alg:sec} or Algorithm \ref{alg:NR} to compute the hitting time $t_h$.
\item[Step 5] Use Eq.\eqref{eq:hamilton_sol_1}-Eq.\eqref{eq:hamilton_sol_bounc} (Eq.\eqref{eq:v_approx} and Eq.\eqref{eq:v_approx2} are excluded if a Newton-Raphson\\ approach is used in Step 4) to propose another state $\vect u^{**}$, \\
\eIf{$\vect u^{**}\in\mathcal{F}_j$ }{
	\begin{addmargin}[1em]{1em}
		accept the proposal.
	\end{addmargin}
        }{
        \begin{addmargin}[1em]{1em}
		set $\vect u^{**}$ to $\vect u_{init}$. 
	\end{addmargin} 
        }
\end{description}
 \caption{BB-HMC to sample from $\varphi(\vect u|\mathcal{F}_j)$}
\end{algorithm}
Note that the transition matrix of BB-HMC satisfies the detailed balance if $t_h$ and $\vect v$ are solved analytically. This is because in that case: i. the mechanism of proposing new state is symmetric; and ii. the bouncing is Hamiltonian preserving. However, if $t_h$ and $\vect v$ are solved numerically or approximately as described in Algorithm \ref{alg:sec} or Algorithm \ref{alg:NR}, although the Hamiltonian can still be preserved, the HMC proposals are no longer strictly symmetrical. To see this, consider one attempts to trace the trajectory from $\left(\vect u(t), \vect p(t)\right)$ to $\left(\vect u(t+t_f),\vect p(t+t_f)\right)$ backwards, starting from $\left(\vect u(t+t_f ),-\vect p(t+t_f)\right)$ (note a momentum negation is applied), due to the approximations in determining the hitting time $t_h^-$ and direction vector $\vect v^-$ for the backward trajectory, it is not guaranteed that equations $t_h^-+t_h^+=t_f$ and $\vect v^-=\vect v^+$, where $t_h^+$ and $\vect v^+$ denote the hitting time and direction vector for the forward trajectory, respectively, can be met exactly. In consequence, for the backward trajectory the system is not guaranteed to reach state $\left(\vect u(t),-\vect p(t)\right)$ after the same amount of time $t_f$. Thus it can be concluded that Algorithm 6 is approximate in its nature, and the accuracy depends on the accuracy of $t_h$ and $\vect v$ used in the algorithm. However, it will be seen in a series of examples that despite the approximation in the BB-HMC algorithm, the BB-HMC based Subset Simulation can still be accurate. 
\subsection{Hamiltonian Monte Carlo method for non-Gaussian distributions}
\noindent
This section focuses on HMC-SS for non-Gaussian distributions. Provided the introductions in Section \ref{sec:HMC_1}-\ref{sec:HMC_3}, with simple modifications, the aforementioned HMC methods can be used to sample from a generic continuous distribution, as long as the gradient of the target PDF (with an unknown normalizing constant) exists. Given these conditions, this framework is particularly suitable to perform reliability analysis in the original probability space without the need to transform it into the standard normal space. 

Consider the conditional PDF 
$\pi(\vect q|\mathcal{F}_j)$, where $\mathcal{F}_j$ is the $j$-th intermediate failure domain in Subset Simulation, and the joint PDF of non-Gaussian variables $\vect q$, $\pi (\vect q)$, is continuously differentiable. Using ideas of HMC, auxiliary multivariate normal momentum variables, $\vect p\sim N(\vect 0, \mathcal{M})$, are introduced so that the Hamiltonian can be written as
\begin{equation}
       \begin{split}
	H(\vect u,\vect p)&=V(\vect u)+K(\vect p)=-\log\pi(\vect q|\mathcal{F}_j)+\frac{\vect p\cdot\mathcal{M}^{-1}\vect p}{2},\\
	 			  &=-\log\pi(\vect q)+\frac{\vect p\cdot\mathcal{M}^{-1}\vect p}{2}-\log I_{\mathcal{F}_j}(\vect q) + \text{const}.\\
	\end{split}\label{eq:HMC_NG}
\end{equation} 
where $I_{F_j} (\vect q)$ is a binary indicator function that gives ``1" if $\vect q\in \mathcal{F}_j$, and gives ``0" otherwise. The last line of Eq. \eqref{eq:HMC_NG} is derived using the property
\begin{equation}
        \pi(\vect q|\mathcal{F}_j)=\frac{\pi(\vect q)I_{\mathcal{F}_j}(\vect q)}{\int_{\vect q\in \mathcal{F}_j}\pi(\vect q)d\vect q}
        \label{eq:HMC_prop}
\end{equation} 
where the denominator, although being unknown, is a constant that can be dropped in HMC. It is seen that Eq.\eqref{eq:HMC_NG} has the same form as Eq.\eqref{eq:hamiltonian_sn}, thus the aforementioned rejection sampling and barrier bouncing techniques can still be used to sample from $\pi(\vect q|\mathcal{F}_j)$. The only difference is that one may need a numerical integration technique to solve the Hamilton's equations. The well-known leapfrog method \cite{verlet1967computer} works as follows to approximately solve Hamilton's equations:
\begin{equation}
       \begin{split}
	\vect p \left( t+\frac{\Delta t}{2}\right)&=\vect p (t)-\frac{\Delta t}{2}\frac{\partial V\left(\vect q(t)\right)}{\partial\vect q(t)},\\
	\vect q \left( t+\Delta t \right)&=\vect q (t)-\Delta t \mathcal{M}^{-1}\vect p (t+\frac{\Delta t}{2}),\\
	\vect p \left( t+ \Delta t \right)&=\vect p \left(t+\frac{\Delta t}{2}\right)-\frac{\Delta t}{2}\frac{\partial V\left(\vect q(t+\Delta t)\right)}{\partial\vect q(t)},
	\end{split}
	\label{eq:lf}
\end{equation} 
where $\Delta t$ is a specified incremental time step. Since the leapfrog method is reversible and it conserves the sympletic structure of the phase space, it is an ideal numerical tool for constructing a HMC method. 

As aforementioned, if a numerical integration is used to solve the Hamilton's equation, the invariance of the Hamiltonian could be violated, consequently the detailed balance of the algorithm could break. Thus, the following Metropolis accept-reject rule is introduced: 
\begin{equation}
\begin{split}
     &\min\left[1,\exp(-H\left(\vect{q}^*,\vect{p}^*)+H(\vect{q},\vect{p})\right)\right]=\\
     &\min\left[1,\exp\left(-V(\vect q^*,\vect p^*)+V(\vect q,\vect p)-K(\vect q^*,\vect p^*)+K(\vect q,\vect p)\right)\right].
\end{split}
	\label{eq:lfcorr}
\end{equation} 
Using the leapfrog method to determine the state of the Hamiltonian system at $t_f$, combined with a Metropolis accept-reject rule in Eq.\eqref{eq:lfcorr}, Algorithm \ref{alg:HMC_rej} and Algorithm \ref{alg:HMCBB} can be adapted to sample from $\pi(\vect q |\mathcal{F}_j)$. Specifically, the modified rejection sampling based HMC to sample from non-Gaussian conditional distribution $\pi(\vect q |\mathcal{F}_j)$ is described as in Algorithm \ref{alg:HMC_nG}.
\begin{algorithm}[!b]
 \SetAlgoNoLine
\label{alg:HMC_nG}
\begin{description}
\item[Step 1] Generate a random initial momentum $\vect p_{init}$ according to $N(\vect 0,\mathcal{M})$, ($\mathcal{M}=I$ is used in this paper). 
\item[Step 2] . Using momentum $\vect p_{init}$ and position $\vect q_{init}$ of a seed sample as initial conditions,  \\ propose a new state $(\vect q^*,\vect p^* )$ via Eq.\eqref{eq:lf} iterated for $L=\text{round}(t_f/\Delta t)$ steps, where $\text{round}(\cdot)$ is the nearest integer function. 
\item[Step 3] Acceptance criteria:\\

\eIf{ $\vect q^*\in\mathcal{F}_j \wedge$ \upshape{rand}$<\min[1,\exp(-H(\vect q^*,\vect p^*)+H(\vect q_{init},\vect p_{init})],$ \upshape{where $\text{rand}\sim U([0,1])$,\\ i.e. the standard uniform  distribution} }{
	\begin{addmargin}[1em]{1em}
		accept the proposal 
	\end{addmargin}
        }{
        \begin{addmargin}[1em]{1em}
		set $\vect q^*$ to $\vect q_{init}.$
	\end{addmargin} 
        }
\end{description}
 \caption{RS-HMC to sample from non-Gaussian distribution $\pi(\vect q|\mathcal{F}_j)$}
\end{algorithm}

\begin{figure}[h]
    \centering
    \includegraphics[width=0.60\textwidth]{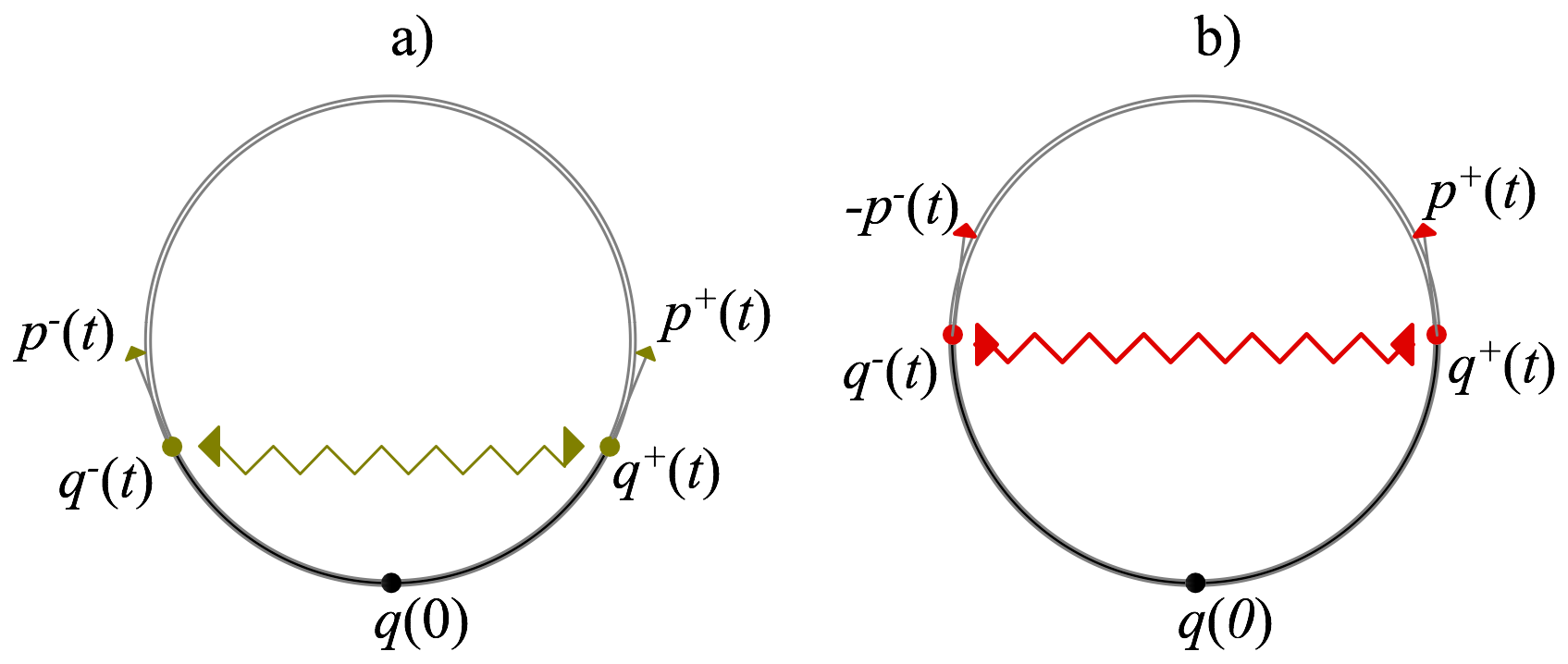}
    \caption{Principles of Algorithm \ref{alg:mean_T}, a) when $\left[\vect p_+^{(i)^T}\cdot\left( \vect q_+^{(i)}-\vect q_-^{(i)}\right) \ge 0 \right]^{(L)}\wedge \left[\vect p_+^{(i)^T}\cdot\left( \vect q_+^{(i)}-\vect q_-^{(i)}\right)\ge 0\right]^{(L)}$ the trajectory of $\vect q(t)$ expands forward and backward along equi-Hamiltonian orbits; b) when $\left[\vect p_+^{(i)^T}\cdot\left( \vect q_+^{(i)}-\vect q_-^{(i)}\right) < 0 \right]^{(L)}\wedge \left[\vect p_+^{(i)^T}\cdot\left( \vect q_+^{(i)}-\vect q_-^{(i)}\right)<0\right]^{(L)}$ the trajectory of $\vect q(t)$ contracts and approaching neighbourhoods that are potentially already explored.
    }
    \label{fig:U_turn}
\end{figure}

The $\Delta t$ can be chosen to be sufficiently small such that it introduces negligible error to the Hamiltonian (i.e., $\exp(-H(\vect q^*,\vect p^* )+H(\vect q_{init},\vect p_{init} ))\approx 1$). Note that in practice the main computational effort in Subset Simulation is usually the evaluation of limit-state functions, and each leapfrog step (except the last step) does not involve limit-state function evaluation, thus the additional cost introduced by using a relatively small $\Delta t$ is often negligible. 
	The $t_f$ in Step 2 of Algorithm \ref{alg:HMC_nG} can be determined adaptively using principles of Algorithm \ref{alg:adaptrule}. In the standard normal space, the periodic structure of the $(\vect q,\vect p)$ orbits is clearly given by the isotropic nature of the space. However, for non-Gaussian spaces the periodic structure is not trivial or unique; in fact, it depends upon the trajectory of $\left(\vect q(t),\vect p(t)\right)$. To overcome this obstacle, a surrogate mean period, here denoted with $\bar T$, is devised using ideas of the No-U-Turn HMC method \cite{hoffman2014no}, and described as in Algorithm  \ref{alg:mean_T}.  The principles of Algorithm \ref{alg:mean_T} are illustrated in Figure \ref{fig:U_turn}. In specific, given an initial position $\vect q(0)$ and initial momentum $\vect p(0)$, two fictitious particles move backward $(\vect q^- (t),\vect p^- (t))$ and forward $(\vect q^+ (t),\vect p^+ (t))$ along an equi-Hamiltonian orbit. At the beginning, their distance expands (green spring in Figure \ref{fig:U_turn}.a)) leading to an optimal exploration of the probability space. The stopping criterion acts when further exploration of the orbit leads on a contraction of the relative distance between the fictitious particles (red spring in Figure \ref{fig:U_turn}.b)). Note that Algorithm \ref{alg:mean_T} does not involve limit-state function evaluations. 
\begin{algorithm}[!b]
  \SetAlgoNoLine
\label{alg:mean_T}
\textbf{Step 1}.For each sample of $N_a$ chains in the Subset Simulation iteration, denoted as \\ \ \ \ \ \ \ \ \ $ (\vect q^{(i)},\vect p^{(i)} )$,  $i=1,2,...,N_a (1/p_0 )$, compute the period of each $\left (\vect q^{(i)},\vect p^{(i)}\right )$,  denoted \\ \ \ \ \ \ \ \ \ as $T^{(i)}$, via:  \\
\begin{addmargin}[1em]{2em}
\ForEach{$L=1,2, \dots$} {
	\begin{addmargin}[1em]{2em}
		forward, evaluate: 
		\begin{equation}
		      \left (\vect q^{(i)},\vect p^{(i)}\right )^{(0)}\xrightarrow{\text{$L$ leapfrog steps Eq.\eqref{eq:lf}}}\left (\vect q_+^{(i)},\vect p_+^{(i)}\right )^{(L)};\nonumber
		\end{equation}
		backward, evaluate:  
		\begin{equation}
		      \left (\vect q^{(i)},-\vect p^{(i)}\right )^{(0)}\xrightarrow{\text{$L$ leapfrog steps Eq.\eqref{eq:lf}}}\left (\vect q_-^{(i)},\vect p_-^{(i)}\right )^{(L)};\nonumber
		\end{equation}	
    	\If{$\left[\vect p_+^{(i)^T}\cdot\left( \vect q_+^{(i)}-\vect q_-^{(i)}\right) < 0 \right]^{(L)}\wedge \left[\vect p_+^{(i)^T}\cdot\left( \vect q_+^{(i)}-\vect q_-^{(i)}\right)<0\right]^{(L)}$}{\begin{addmargin}[1em]{1em}
		 set $T^{(i)}=2L\Delta t$\\
		 break\;
		  \end{addmargin}
    		}\end{addmargin}
	 }
\end{addmargin}
\textbf{Step 2} Compute 
\begin{equation}
	\bar T = \frac{p_0}{N_a}\sum_{i=1}^{\frac{N_a}{p_0}} T^{(i)}\nonumber	      
\end{equation}
 \caption{Estimate the mean period $\bar T$}
\end{algorithm}
%
%
%

Provided the mean period $\bar T$, the adaptive rule to select $t_f$ follows Algorithm \ref{alg:adaptrule} and leads to Algorithm 9.  \\
\begin{algorithm}[H]
  \SetAlgoNoLine
\label{alg:Adap_rule_tf}
\begin{description}
\item[Step 1] For the current intermediate step $j$ of Subset Simulation,\\
\eIf{$j=1$}{\begin{addmargin}[1em]{1em}
		 initialize $t_f$ as, e.g., $t_f=\bar T_0/8$, where $\bar T_0/8$ represents the mean period of the Hamiltonian system\\
		
		  \end{addmargin}
    		}
		{\begin{addmargin}[1em]{2em}
    		  initialize $t_f$ as the $t_f$ selected in step $j-1$;\
   		 \end{addmargin}}
\item[Step 2]   Compute the acceptance rate, denoted as $a$, for every $N_a$ chains simulated. If \\ $a<a_{low}$ (it is suggested $a_{low}=0.3)$, set $t_f\leftarrow(\bar T/2\pi)\sin^{-1}\{\sin(\bar T/2\pi)\exp[(a-a_{low})/2]\}$,\\ where $\bar T$ is the mean period for samples in each $N_a$ chain. Similarly, if $a>a_{up}$ (it is suggested $a_{up}=0.5$), set $t_f\leftarrow(\bar T/2\pi)\sin^{-1}\{\sin(t_f)\exp[(a-a_{up})/2]\}$.  
\end{description}
 \caption{Adaptive rule to select $t_f$}
\end{algorithm}
$\ $\\
The mean period $\bar T_0$ in Step 1 of Algorithm \ref{alg:Adap_rule_tf} can be estimated using Algorithm \ref{alg:mean_T} with randomly drawn $(\vect q^{(i)},\vect p^{(i)})$ samples of $\pi(\vect q, \vect p)$ at the beginning of Subset Simulation. The estimation of $\bar T_0$ also does not involve limit-state function evaluations. 

The barrier bouncing based HMC to sample from non-Gaussian conditional distribution $\pi(\vect q |\mathcal{F}_j)$ is described in Algorithm \ref{alg:HMCBB_ng}.
\begin{algorithm}[h!]
  \SetAlgoNoLine
\label{alg:HMCBB_ng}
\begin{description}
\item[Step 1] Generate a random initial momentum $\vect p_{init}$ according to $N(\vect 0,\mathcal{M})$, ($\mathcal{M}=I$ is used in this paper). 
\item[Step 2] Using momentum $\vect p_{init}$ and position $\vect u_{init}$ of a seed sample as initial conditions,  \\ propose a new state $(\vect u^*,\vect p^* )$ via Eq.\eqref{eq:lf} iterated for $L=\text{round}(t_f/\Delta t)$ steps. 
\item[Step 3] Acceptance criteria:
\eIf{ $\vect u^*\in\mathcal{F}_j\wedge$ \upshape{rand}$ <\min\left[1,\exp\left(-H(\vect q^*,\vect p^*)+H(\vect q_{init},\vect p_{init})\right)\right]$\\}{
	\begin{addmargin}[1em]{1em}
		accept the proposal 
	\end{addmargin}
        }{
        \begin{addmargin}[1em]{1em}
		go to step 4
	\end{addmargin} 
        }
\item[Step 4] Use $(\vect u^*,\vect p^* )$ and $(\vect u_{init},\vect p_{init} )$ and their corresponding limit-state function values as initials of Algorithm \ref{alg:sec} or Algorithm \ref{alg:NR} to compute the hitting time $t_h$.
\item[Step 5] Leapfrog round$(t_h/\Delta t)$ steps from state $(\vect q_{init},\vect p_{init} )$ to obtain state $(\vect q_b,\vect p_b)$. Use \\ Eq.\ref{eq:hamilton_sol_2} to compute $(\vect q_a, \vect p_a)$ if Newton-Raphson algorithm is used in Step 4, else use Eq.\eqref{eq:hamilton_sol_2}-\eqref{eq:v_approx2} to compute $(\vect q_a,\vect p_a)$. Leapfrog round$[(t-t_h)/\Delta t]$ steps from state $(\vect q_a,\vect p_a )$\\ to obtain a proposed state $(\vect q^{**},\vect p^{**} )$:

\eIf{ $\vect u^{**}\in\mathcal{F}_j\wedge$ \upshape{rand}$ <\min\left[1,\exp\left(-H(\vect q^{**},\vect p^{**})+H(\vect q_{init},\vect p_{init})\right)\right]$}{
	\begin{addmargin}[1em]{1em}
		accept the proposal 
	\end{addmargin}
        }{
        \begin{addmargin}[1em]{1em}
		set $\vect q^{**}$ to $\vect q_{init}$
	\end{addmargin} 
        }
\end{description}
 \caption{BB-HMC to sample from non-Gaussian distribution $\pi(\vect u|\mathcal{F}_j)$}
\end{algorithm}
Similarly, the $t_f$ in Step 2 of Algorithm \ref{alg:HMCBB_ng} can be determined adaptively using Algorithms \ref{alg:mean_T} and \ref{alg:Adap_rule_tf}. Specifically, the rule to adapt $t_f$ is given as: compute the acceptance rate a for each $N_a$ chains simulated, if $a<a^*$ (it is suggested $a^*=0.8$), set $t_f\leftarrow(\bar T/2\pi)\sin^{-1}\{\sin(2\pi t_f/\bar{T})\exp[(a-a^*)/2]\}$, where the mean period $\bar T$ is obtained from Algorithm \ref{alg:Adap_rule_tf}. 
%
%
%
\break 
\section{Numerical Investigations}\label{sec:NI}
\subsection{Behavior of HMC}\label{sec:B_HMC}
\subsubsection{Bivariate standard normal distribution }
\begin{figure}[t]
    \centering
    \includegraphics[width=0.82\textwidth]{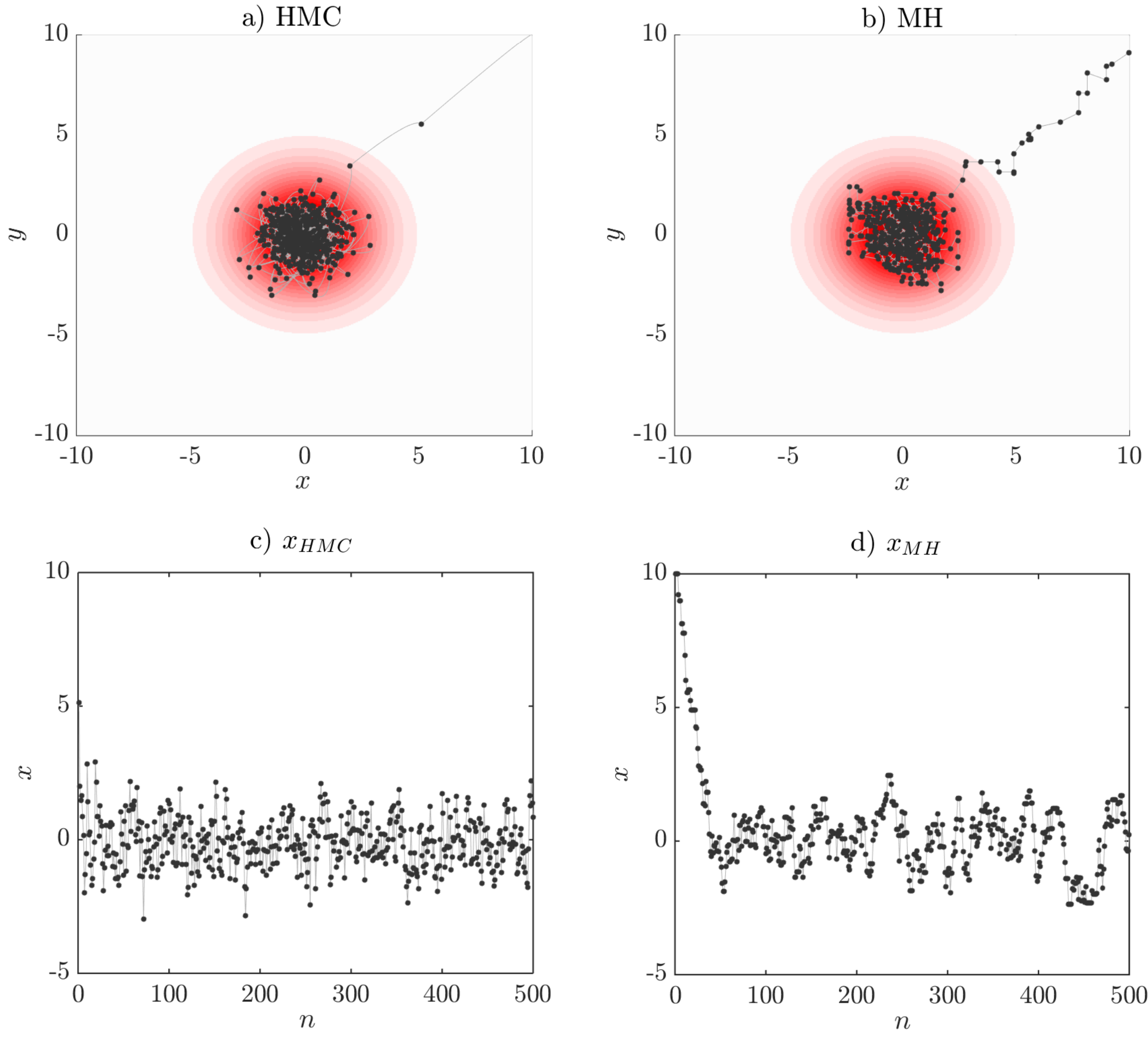}
    \caption{Trajectories obtained from a) HMC algorithm, and b) MH-CW algorithm. First coordinate values for trajectories from c) HMC algorithms d) MH-CW algorithm.
    }
    \label{fig:HMC_2}
\end{figure}

\noindent
In this section, a series of simple examples are introduced to aid an intuitive understanding of how HMC operates. In the first example, we use HMC to sample from a bivariate standard normal distribution, starting with a seed $\vect u_0=(10,10)$ at the far tail region of the distribution. Parameters of HMC are set as $t_f=\pi/3$, $\alpha = 0$ (see Eq.\eqref{eq:pr}). The trajectory of 500 HMC iterations are shown in Figure \ref{fig:HMC_2} a). 

\begin{figure}[!h]
    \centering
    \includegraphics[width=0.82\textwidth]{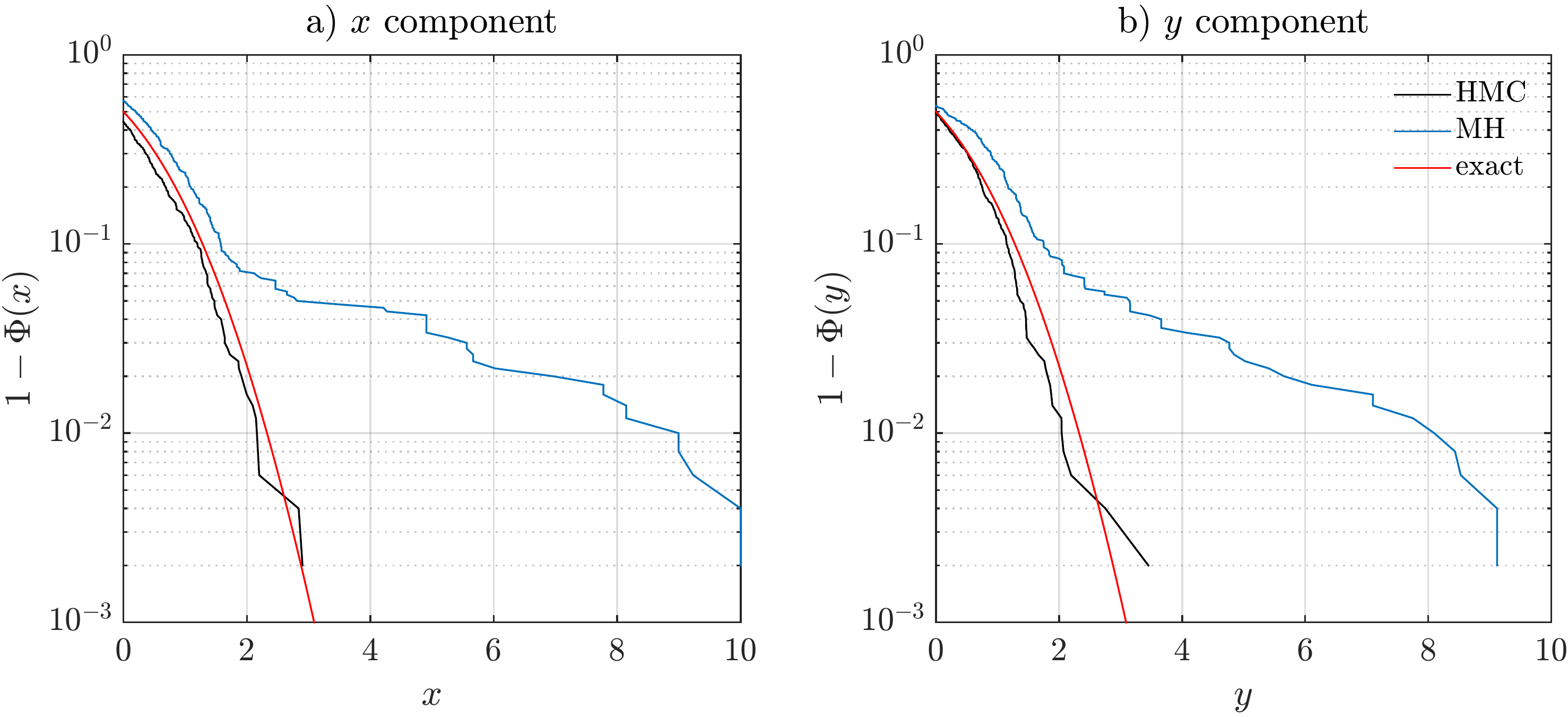}
    \caption{Marginal complementary CDFs obtained from CW-MH and HMC.
    }
    \label{fig:HMC_3}
\end{figure}

For comparison, the trajectory of 500 iterations of the component-wise Metropolis Hastings (CW-MH) algorithm with a uniform transition distribution of width 2 are shown in Figure  \ref{fig:HMC_2} b). Moreover, the first coordinate values of the samples in Figure \ref{fig:HMC_2} a-b) are plotted in Figure \ref{fig:HMC_2} c-d) against the number of iterations respectively. The marginal sample complementary CDF associated with two coordinates obtained from HMC and CW-MH, compared with the analytical solution are shown in Figure \ref{fig:HMC_3}.

In Figure \ref{fig:HMC_2}-\ref{fig:HMC_3}, it is clearly seen that: i. HMC marches to the high probability density region of the standard normal distribution faster than CW-MH; ii. the autocorrelation of HMC samples is noticeably lower than that of the CW-MH samples; and iii. HMC leads to a more effective estimate of the target CDF or CCDF than CW-MH. 

Point i. is not of great significance in the context of Subset Simulation because the starting points of the subset chains are already distributed accordingly the target density (a property named perfect sampling \cite{au2012discussion}); however, point ii. and iii. are rather important. In fact, point ii. would result in a significantly decrease of the coefficient of variation (c.o.v) of the failure probability estimate, which depends on the autocorrelation lag of the chain, and point iii. would result in a reduction of the error related to the conditional failure probability of each intermediate failure domain, and therefore reduces the bias of the failure probability estimate. 

\begin{figure}[t]
    \centering
    \includegraphics[width=0.82\textwidth]{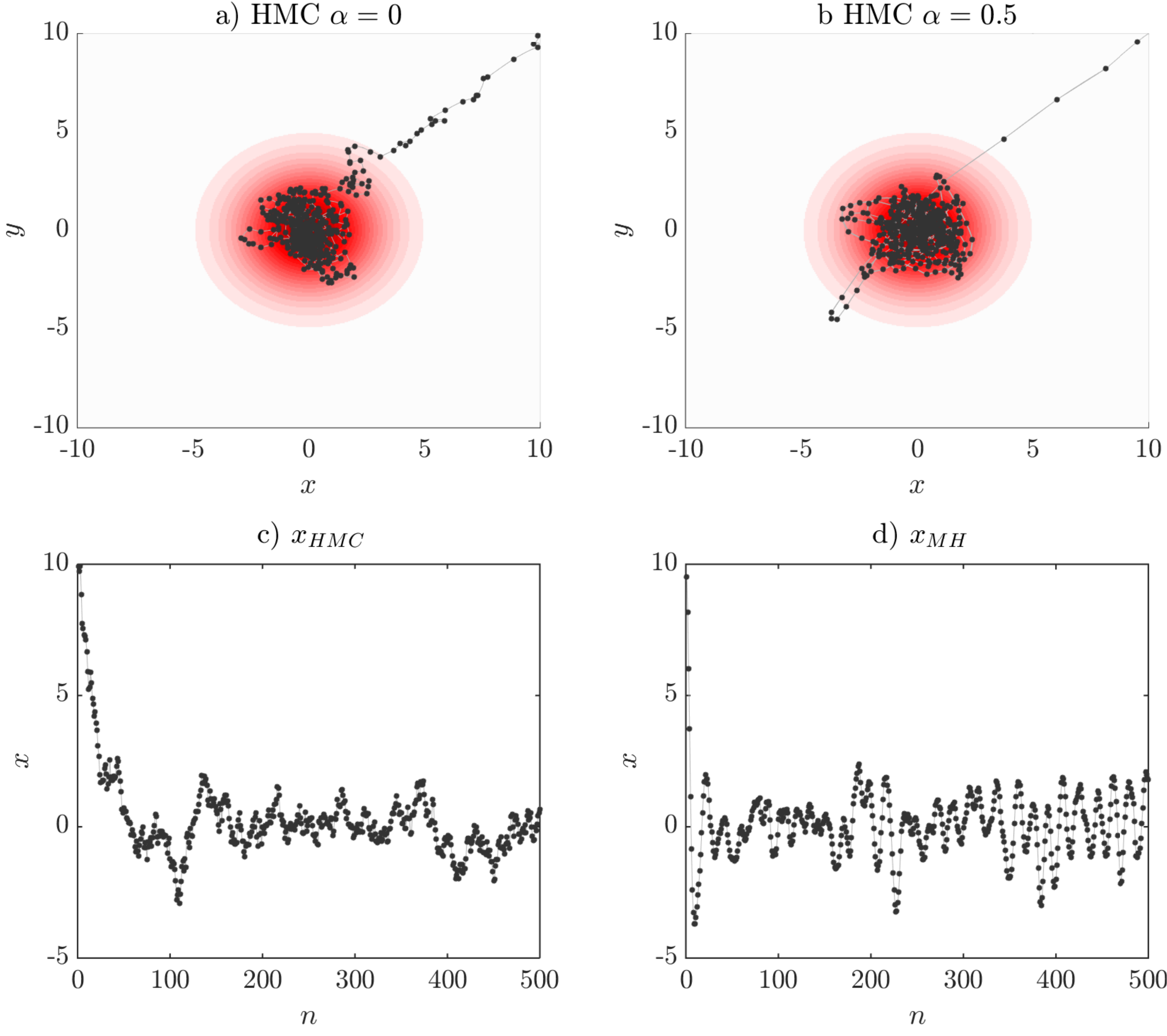}
    \caption{Trajectories obtained from HMC algorithms with different $\alpha$s; a) $\alpha=0$, b) $\alpha=0.5$; and first coordinate values for trajectories from HMC algorithms with different c)$\alpha=0$, d) $\alpha=0.5$.
    }
    \label{fig:HMC_4}
\end{figure}

\begin{figure}[!h]
    \centering
    \includegraphics[width=0.82\textwidth]{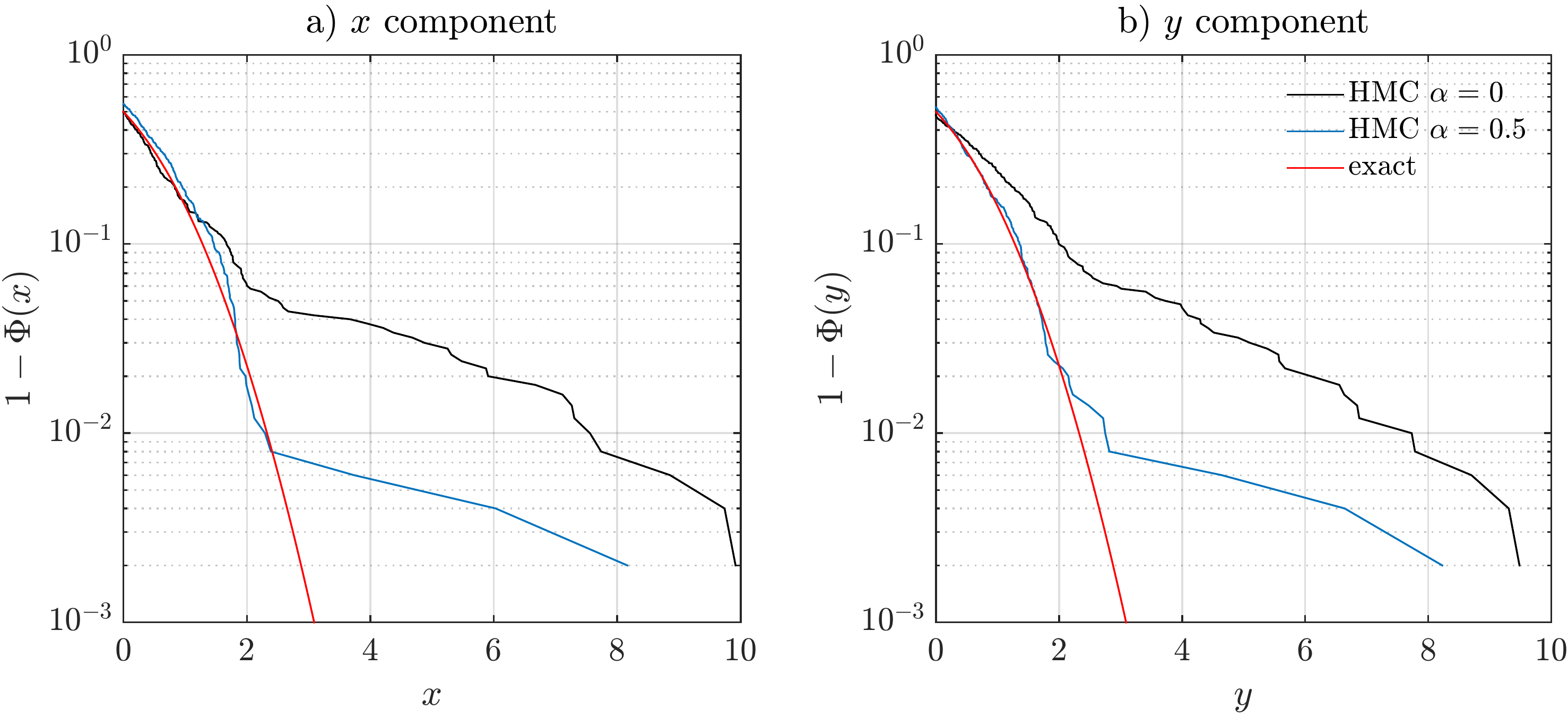}
    \caption{Marginal complementary CDFs obtained from HMC algorithms with different $\alpha$s.
    }
    \label{fig:HMC_5}
\end{figure}

Note that the aforementioned observation is based on a good parameter setting of HMC. If the parameters of HMC are set poorly, HMC would display a more server random walk behavior. However, even for a poor setting of $t_f$, the random walk behavior of HMC can still to some extent be suppressed by the partial momentum refreshment technique. To illustrate this idea, Figure \ref{fig:HMC_4} a) shows the trajectory of 500 HMC iterations using $t_f = \pi/10, \alpha=0$, while Figure \ref{fig:HMC_4}  b) shows the trajectory using $t_f = \pi/10, \alpha=0.5$. The first coordinate values corresponding to trajectories in Figure \ref{fig:HMC_4}  a-b) are plotted in Figure  \ref{fig:HMC_4}  c-d). The marginal complementary CDF associated with two coordinates obtained from HMC using $\alpha=0$ and $\alpha=0.5$, compared with the analytical solution are shown in Figure \ref{fig:HMC_5}.

\begin{figure}[t]
    \centering
    \includegraphics[width=0.82\textwidth]{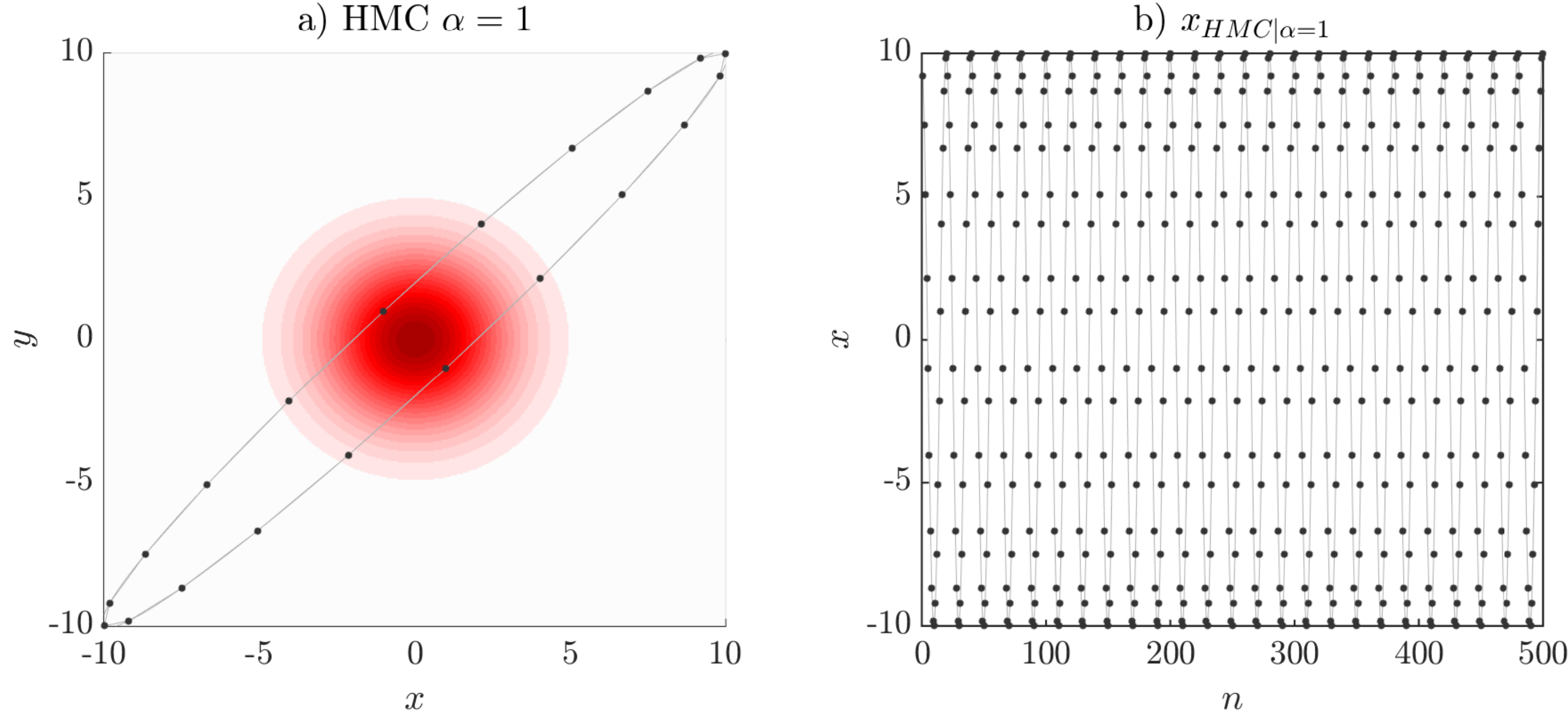}
    \caption{Trajectory and first coordinate values from HMC algorithm using $\alpha=1$.
    }
    \label{fig:HMC_6}
\end{figure}

In Figure \ref{fig:HMC_4}-\ref{fig:HMC_5}, it is seen that the partial momentum refreshment technique suppresses the random walk behavior and leads to a more effective exploration of the probability space. However, one should use this technique with care because in the limiting case it can break the ergodicity of the Markov chain. To see the limiting behavior of the partial momentum refreshment technique, Figure 6 a) shows the trajectory of 500 HMC iterations using $t_f = \pi/10$, $\alpha = 1$, $\vect q_0 =[10, 10],\ \vect p_0=[1,-1]$, and Figure \ref{fig:HMC_6} b) shows the corresponding first coordinate values. As indicated by Eq.\eqref{eq:hamilton_sol}, the trajectory in Figure \ref{fig:HMC_6} is an ellipse. Therefore, the chain is periodic (i.e. trapped in an elliptical orbit) and it will not explore effectively the probability space. Periodic Markov chains are not ergodic and therefore they cannot be used for statistical computing purposes. 

\subsubsection{Truncated bivariate standard normal distribution}
\noindent
Since constraint is not involved in the previous examples, no rejection or barrier bouncing technique is needed for HMC. To study the behavior of rejection sampling based HMC (RS-HMC) and barrier bouncing based HMC (BB-HMC), now the two HMC approaches are used to sample from a truncated bivariate standard normal distribution with constraint $2\sqrt{2}-u_1-u_2\le 0$, starting with a seed $\vect u_0=(10,10)$. Parameters of the two HMC approaches are set as $t_f=\pi/5$, $\alpha=0$. 

\begin{figure}[!h]
    \centering
    \includegraphics[width=0.88\textwidth]{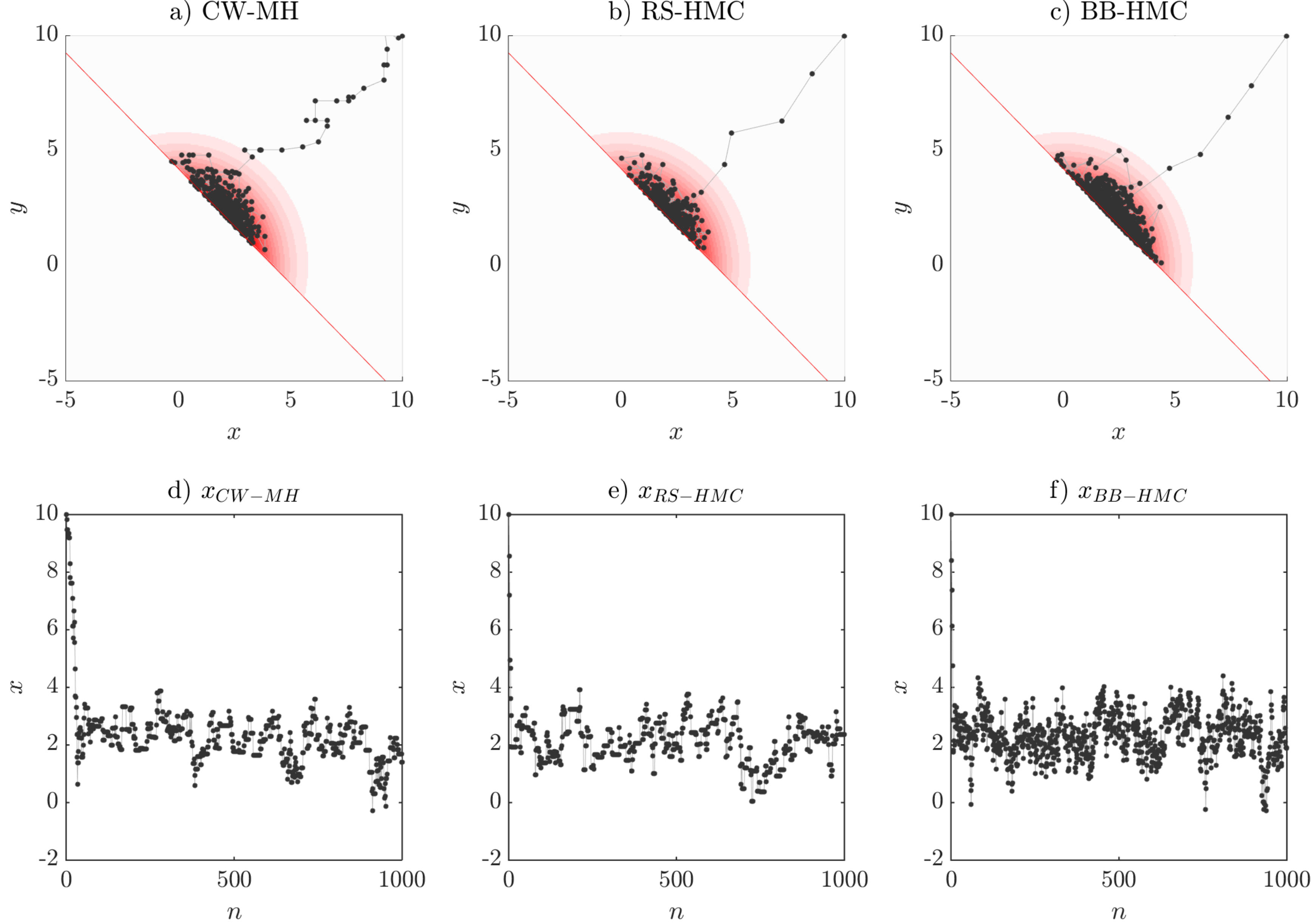}
    \caption{Top: trajectories obtained from a) CW-MH, b)RS-HMC, and c) BB-HMC algorithms. Bottom: first coordinate values for trajectories from d) CW-MH, e) RS-HMC, and f) BB-HMC algorithms.
    }
    \label{fig:HMC_7}
\end{figure}

The trajectory of 1,000 iterations of the CW-MH algorithm with a uniform transition distribution of width 2 are shown in Figure \ref{fig:HMC_7} a). The trajectory of 1,000 RS/BB-HMC iterations are shown in Figure \ref{fig:HMC_7} b) and c). The first coordinate values of the samples for the three cases are plotted in Figure \ref{fig:HMC_7} d), e), and f).
Similar to previous observations on HMC, it is seen in Figure  \ref{fig:HMC_7}  that the two HMC approaches display less random walk behavior, and the BB-HMC is especially effective (at the cost of additional computations in simulating the bouncing process). 
\break
\subsubsection{Banana-shaped bivariate non-Gaussian distribution}
\noindent
Next, consider a ``banana-shaped'' PDF derived as follow. Given two correlated Gaussian random variables, $u_1$ and $u_2$, with zero mean, unit variance, and correlation coefficient $\rho$, consider the following transformation:
\begin{equation}
\begin{split}
x&=u_1a, \\
y&=\frac{u_2}{a}+b(u_1^2+a^2),
\end{split}
\end{equation}
where $a\in \mathbb{R}$ and $b\in \mathbb{R}$. Then, the joint PDF of $(x,y)$ can be written as 
\begin{equation}
\begin{split}
f_{XY}(x,y) = &\frac{1}{Z}\exp\left[-\frac{1}{2(1-\rho^2)}\left( \frac{x^2}{a^2}+a^2\left(y-b\frac{x^2}{a^2}-ba^2\right)^2\right.\right.\\
&\left.\left. -2\rho\left(y-b\frac{x^2}{a^2}-ba^2\right)\right)\right],
\end{split}
\end{equation}
where $Z$ is a normalizing constant. In this example, we set $a=1.15$, $b= 0.5$, $\rho=0.9$, Figure 8 a) shows the contour plot of the ``banana-shaped'' distribution. 

\begin{figure}[!b]
    \centering
    \includegraphics[width=0.9\textwidth]{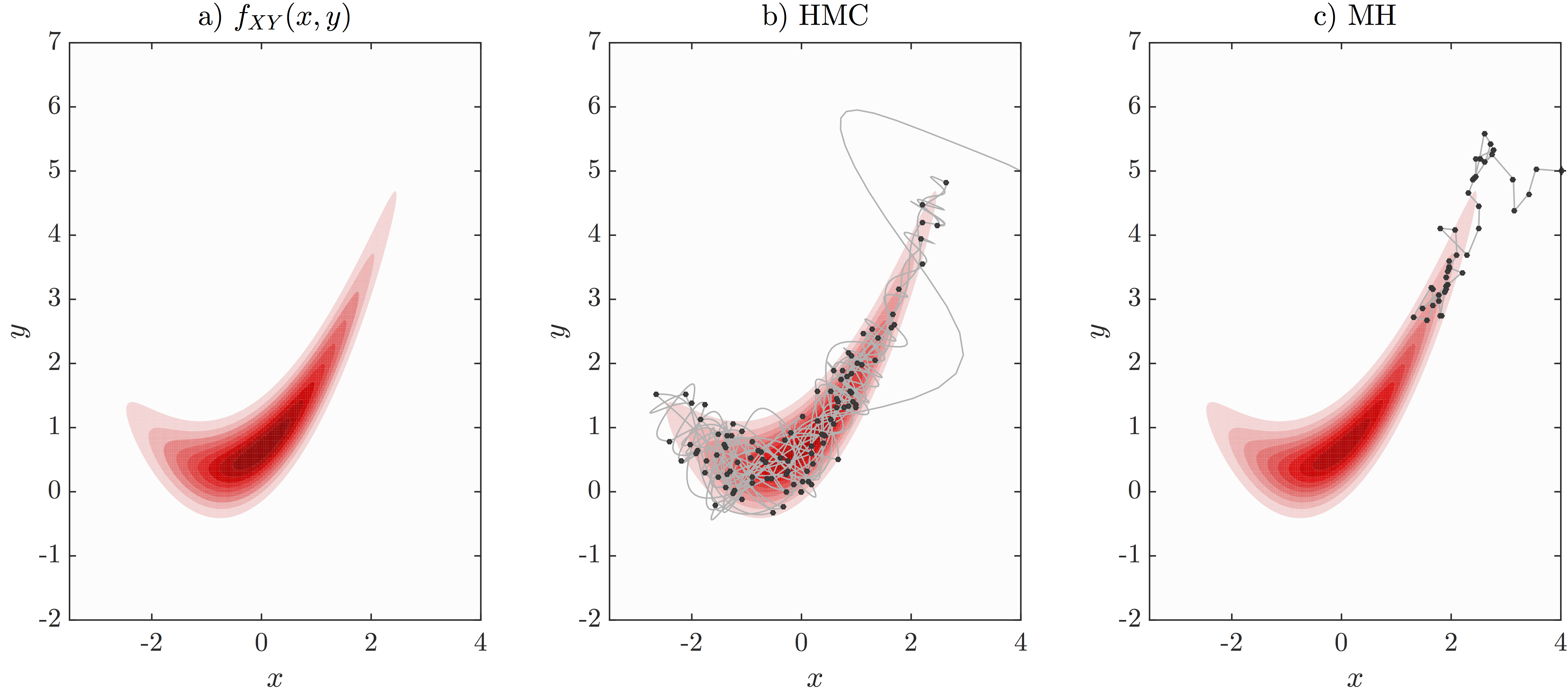}
    \caption{a) PDF Contour b)-c) Trajectories obtained from HMC and MH algorithms.
    }
    \label{fig:HMC_8}
\end{figure}

We use leapfrog based HMC to sample from the banana-shaped distribution, starting with a seed $\vect q_0=(4,5)$ at the far tail region of the distribution. Parameters of HMC are set as $\Delta t=0.05$,  $L=\text{round}(t_f/\Delta t)=\text{round}(\pi/3/0.05)=21$. The trajectory of 100 HMC iterations are shown in Figure \ref{fig:HMC_8} b). For comparison, the trajectory of 100 iterations of the Metropolis Hastings algorithm using a uniform transition distribution within a square of width 1 are shown in  Figure \ref{fig:HMC_8} c). One can observe the efficiency of HMC in reaching the bulk of the probability density in only one iteration. This example shows that HMC is particularly suitable for probability densities that are ``narrow'' and confined in specific region of the space. This is the typical case of high dimensional spaces, where the bulk of probability lies in specific confined regions named typical set. Conversely for these settings, the proposals of the random walk based Metropolis Hastings algorithm are highly likely to be rejected, thus as it is seen in Figure \ref{fig:HMC_8} c) the effectiveness of Metropolis Hastings algorithm is noticeably lower than the HMC algorithm.
\break
\subsection{Reliability example in standard normal space}
\noindent
In this section, HMC-SS will be tested and compared with the Component-wise Metropolis-Hastings based Subset Simulation (CWMH-SS) for reliability examples formulated in standard normal space. In all the following examples of this section, the parameters of Subset Simulation are chosen as $N=1,000$, where $N$ is the number of samples in each of the intermediate steps, and $p_0=0.1$, where $p_0$ is the percentile for determining the nested failure domains. For the rejection sampling (RS-) and barrier bouncing (BB-) based HMC-SS approaches, the parameter $\alpha$ is set to zero. For the conventional CWMH-SS, a uniform distribution of width 2 is used in the Metropolis Hastings algorithm. For all the examples, Subset Simulation is independently performed 500 times, so that the sample mean and c.o.v of the results can be obtained. Note that despite the noticeable influence the parameter $\alpha$ could have on behavior of HMC algorithm, it is found the effect of $\alpha$ on the performance of HMC-SS is to some extent inconclusive. In this paper the numerical investigations of HMC-SS are performed without considering the issue regarding selection of $\alpha$.
\subsubsection{Reliability example with linear limit-state function}
\noindent
Consider a limit-state surface defined by a linear function
\begin{equation}
G(\vect u) = \beta_0 -\frac{1}{\sqrt{n}}\sum_{i=1}^n\vect u_i,
\label{eq:lin}
\end{equation}
where n is the dimension. Regardless of the dimension $n$, the failure probability of the limit-state function in Eq.\eqref{eq:lin} is $\Phi(-\beta_0)$ in which $\Phi(\cdot)$  is the cumulative distribution function of the standard normal distribution. To study the influence of the level of the failure probability, Subset Simulation is performed for a dimension of $n=100$ and a sequence of $\beta_0$ values. Table 1a and Table 1 b) illustrate the sample mean and coefficient of variation of the failure probabilities obtained from RS-HMC-SS, CWMH-SS, and BB-HMC-SS (implemented with Newton-Raphson and secant). The tables also show the mean number of limit-state function evaluations, denoted as NG, for each method. 
%
\begin{table}[!h]
\renewcommand{\thetable}{\arabic{table}.a)}
\centering
\caption{Performance of HMC-SS for various probability levels.}
\begin{tabular}{*8c}
\toprule
$\beta_0$ &  \multicolumn{3}{c}{RS-HMC} & \multicolumn{3}{c}{CW-MH} & Exact \\ \midrule
  {} & $\hat P_f$   & c.o.v.    & NG   & $\hat P_f$ & c.o.v & NG & $P_f$\\
  2.0 & $2.26\times10^{-2}$ & 0.14 & 1900 & 2.30$\times10^{-2}$ & 0.14 & 1900 & 2.28$\times10^{-2}$\\
  3.0 & $1.34\times10^{-3}$ & 0.25 & 2908 & 1.37$\times10^{-3}$ & 0.27 & 2944 & 1.35$\times10^{-3}$\\
  4.0 & $3.18\times10^{-5}$ & 0.35 & 4600 & 3.27$\times10^{-5}$ & 0.40 & 4600 & 3.17$\times10^{-5}$\\
  5.0 & $2.78\times10^{-7}$ & 0.43 & 6403 & 2.97$\times10^{-7}$ & 0.62 & 6418 & 2.87$\times10^{-7}$\\
  6.0 & $0.98\times10^{-9}$ & 0.52 & 8668 & 1.03$\times10^{-9}$ & 0.68 & 8754 & 0.99$\times10^{-9}$\\
              \bottomrule
\end{tabular}
\label{tab:1a}
\bigskip
\addtocounter{table}{-1}
\renewcommand{\thetable}{\arabic{table}.b)}
\centering
\caption{Performance of HMC-SS for various probability levels.}
\begin{tabular}{*8c}
\toprule
$\beta_0$ &  \multicolumn{3}{c}{BB-HMC (Newton)} & \multicolumn{3}{c}{BB-HMC (Secant)} & Exact \\ \midrule
  {} & $\hat P_f$   & c.o.v.    & NG   & $\hat P_f$ & c.o.v & NG & $P_f$\\
  2.0 & $2.28\times10^{-2}$ & 0.12 & 3427 & 2.28$\times10^{-2}$ & 0.12 & 4250 & 2.28$\times10^{-2}$\\
  3.0 & $1.33\times10^{-3}$ & 0.19 & 7033 & 1.37$\times10^{-3}$ & 0.18 & 8930 & 1.35$\times10^{-3}$\\
  4.0 & $3.20\times10^{-5}$ & 0.25 & 16309 & 3.18$\times10^{-5}$ & 0.24 & 20164 & 3.17$\times10^{-5}$\\
  5.0 & $2.89\times10^{-7}$ & 0.29 & 25414 & 2.83$\times10^{-7}$ & 0.30 & 32352 & 2.87$\times10^{-7}$\\
  6.0 & $1.01\times10^{-9}$ & 0.37 & 37266 & 1.00$\times10^{-9}$ & 0.39 & 47753 & 0.99$\times10^{-9}$\\
              \bottomrule
\end{tabular}
\label{tab:1b}
\end{table}

It is seen from Table \ref{tab:1a} and Table \ref{tab:1b} that the HMC-SS approaches are at least as accurate as the conventional CW-MH-SS approach, while the efficiency of RS-HMC-SS is noticeably higher than CWMH-SS for low probability levels as evidenced by the lower c.o.v's achieved by a similar number of limit-state function evaluations. It is noted that although approximate solutions of $t_h$ and $\vect v$ are used in the BB-HMC algorithms so that the detailed balance does not rigorously hold, the accuracy of BB-HMC-SS using both Newton Raphson and secant method seems intact. The BB-HMC-SS approach achieves lowest c.o.v. for the same number of Subset Simulation runs, but requires a larger number of limit-state function evaluations, which is mainly due to the effort in solving the hitting time via the Newton-Raphson or secant method. As expected, the number of limit-state function evaluations of BB-HMC-SS using Newton-Raphson method is noticeably smaller than the one using secant method, at the cost of gradient computations. 

Compared with the RS-HMC-SS approach, it seems the application of BB-HMC-SS in general reliability problems is not attractive. However, in response surface based reliability analysis where analytical surrogate limit-state function is available, one may find analytical solution for the gradient of the response surface or, ideally, for the hitting time. Consequently, in the BB-HMC-SS algorithm the additional computational cost can be eliminated. To illustrate this idea, Table \ref{tab:2} shows the performance of BB-HMC-SS using the analytical hitting time and directional vector normal to the constraint. As expected, it is seen from Table 2 that the number of limit-state function evaluations of BB-HMC-SS using the analytical hitting time is now similar to the RS-HMC-SS and CWMH-SS approaches, yet the c.o.v achieved by BB-HMC-SS is significantly smaller. 
%
%
\begin{table}[!t]
\centering
\caption{Performance of BB-HMC using analytical hitting times.}
\begin{tabular}{*5c}
\toprule
$\beta_0$ &  \multicolumn{3}{c}{BB-HMC (Analytical Time)} & Exact \\ \midrule
  {} & $\hat P_f$   & c.o.v.    & NG   & $P_f$\\
  2.0 & $2.25\times10^{-2}$ & 0.12 & 1900 & 2.28$\times10^{-2}$\\
  3.0 & $1.34\times10^{-3}$ & 0.17 & 2836 & 1.35$\times10^{-3}$\\
  4.0 & $3.17\times10^{-5}$ & 0.21 & 4600 & 3.17$\times10^{-5}$\\
  5.0 & $2.84\times10^{-7}$ & 0.28 & 6400 & 2.87$\times10^{-7}$\\
  6.0 & $0.99\times10^{-9}$ & 0.35 & 8731 & 0.99$\times10^{-9}$\\
              \bottomrule
\end{tabular}
\label{tab:2}
\end{table}

\begin{figure}[h]
    \centering
    \includegraphics[width=0.51\textwidth]{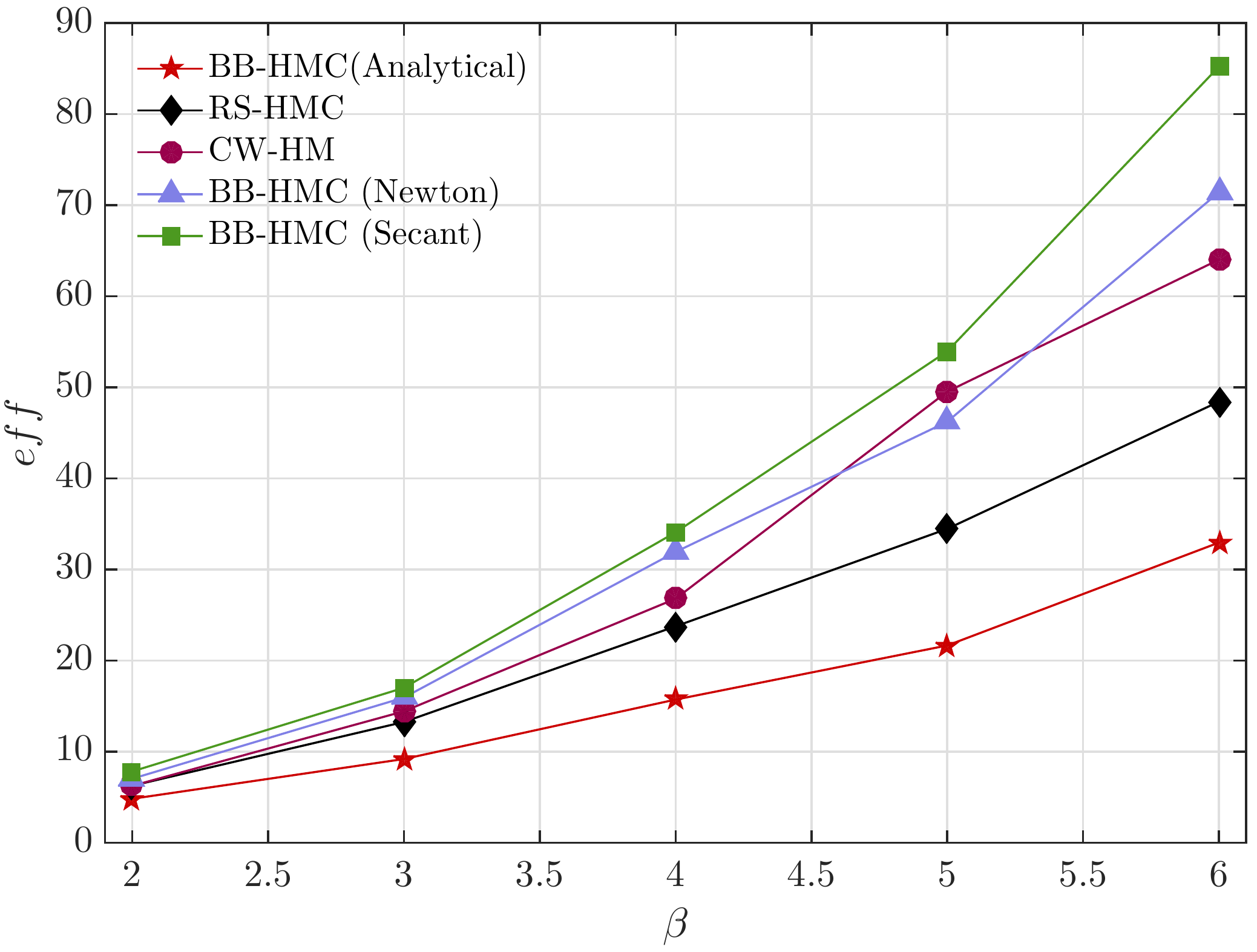}
    \caption{The $\textit{eff}$-$\beta$ curves for various methods.
    }
    \label{fig:HMC_9}
\end{figure}

In the following, it is used $\textit{eff} = \text{c.o.v.} \sqrt{NG}$ \cite{au2007application} as a measure of the efficiency of sampling method (a low $\textit{eff}$ indicates high efficiency). Figure \ref{fig:HMC_9} illustrates the variation of $\textit{eff}$ with the generalized reliability index, expressed by $\beta = -\Phi^{-1}(P_f)$ where $\Phi^{-1}(\cdot)$ is the inverse CDF function of standard normal distribution, for various methods. Note that in this example $\beta = \beta_0$. Figure \ref{fig:HMC_9} suggests similar conclusions for various HMC-SS approaches compared with the CWMH-SS approach as discussed above. 

To study the influence of the dimension, $\beta_0$ is fixed to 4 and Subset Simulation is performed for a sequence of dimensions ranging from 10 to 1000, the results are reported in Table \ref{tab:3a} and Table \ref{tab:3b}. It is seen from Table  \ref{tab:3a} and Table \ref{tab:3b} that the performance of all sampling methods considered here is not sensitive to dimension. 
%
\begin{table}[!h]
\renewcommand{\thetable}{\arabic{table}.a)}
\centering
\caption{Performance of HMC-SS for various dimensions.}
\begin{tabular}{*8c}
\toprule
$n$ &  \multicolumn{3}{c}{RS-HMC} & \multicolumn{3}{c}{CW-MH} & Exact \\ \midrule
  {} & $\hat P_f$   & c.o.v.    & NG   & $\hat P_f$ & c.o.v & NG & $P_f$\\
    10 & $3.12\times10^{-5}$ & 0.32 & 4600 & 3.29$\times10^{-5}$ & 0.39 & 4600 & 3.17$\times10^{-5}$\\
  100 & $3.18\times10^{-5}$ & 0.35 & 4600 & 3.27$\times10^{-5}$ & 0.40 & 4600 & 3.17$\times10^{-5}$\\
  500 & $3.20\times10^{-5}$ & 0.34 & 4600 & 3.22$\times10^{-5}$ & 0.39 & 4600 & 3.17$\times10^{-5}$\\
1000 & $3.19\times10^{-5}$ & 0.33 & 4600 & 3.26$\times10^{-5}$ & 0.39 & 4600 & 3.17$\times10^{-5}$\\
              \bottomrule
\end{tabular}
\label{tab:3a}
\bigskip
\addtocounter{table}{-1}
\renewcommand{\thetable}{\arabic{table}.b)}
\centering
\caption{Performance of HMC-SS for various dimensions.}
\begin{tabular}{*8c}
\toprule
$n$ &  \multicolumn{3}{c}{BB-HMC (Newton)} & \multicolumn{3}{c}{BB-HMC (Secant)} & Exact \\ \midrule
  {} & $\hat P_f$   & c.o.v.    & NG   & $\hat P_f$ & c.o.v & NG & $P_f$\\
    10 & $3.23\times10^{-5}$ & 0.23 & 16557 & 3.11$\times10^{-5}$ & 0.20 & 20242 & 3.17$\times10^{-5}$\\
  100 & $3.20\times10^{-5}$ & 0.25 & 16309 & 3.18$\times10^{-5}$ & 0.24 & 20164 & 3.17$\times10^{-5}$\\
  500 & $3.13\times10^{-5}$ & 0.22 & 16416 & 3.24$\times10^{-5}$ & 0.23 & 20156 & 3.17$\times10^{-5}$\\
1000 & $3.19\times10^{-5}$ & 0.24 & 16340 & 3.17$\times10^{-5}$ & 0.20 & 20215 & 3.17$\times10^{-5}$\\
              \bottomrule
\end{tabular}
\label{tab:3b}
\end{table}

\subsubsection{Reliability example with nonlinear limit-state function}
\begin{figure}[!h]
    \centering
    \includegraphics[width=0.805\textwidth]{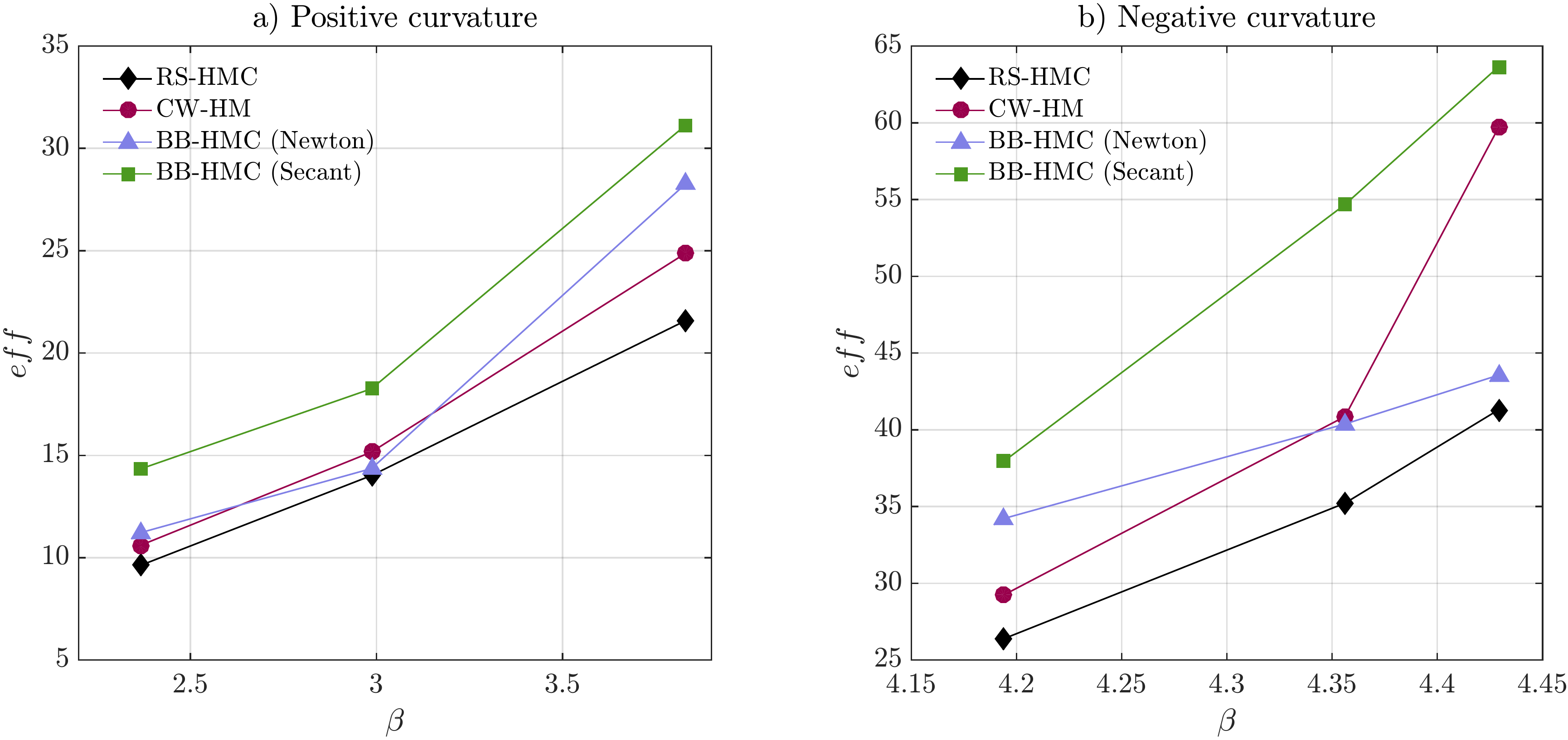}
    \caption{The $\textit{eff}$-$\beta$ curves for various methods and curvature cases; a) positive curvature, b) negative curvature.
    }
    \label{fig:HMC_10}
\end{figure}

\noindent
Consider a nonlinear limit-state function expressed by
\begin{equation}
G(\vect u) = \beta_0 -\frac{1}{\sqrt{n}}\sum_{i=1}^n\vect u_i -\frac{\kappa}{4}(\vect u_1-\vect u_2)^2,
\label{eq:lsf_nl}
\end{equation}
where $n$ is the dimension and $\kappa$ is a curvature parameter to control the nonlinearity of the function. The failure probability of Eq.\eqref{eq:lsf_nl} has an analytical solution and it is independent of the dimension \cite{ditlevsen1996str}. To study the influence of nonlinearity, Subset Simulation is performed for $n=100$, $\beta_0=4.0$ and a sequence of $\kappa$ values. Table \ref{tab:4a} and \ref{tab:4b} show the sample means and coefficients of variation of the failure probabilities obtained from different methods. Figure \ref{fig:HMC_10} a) illustrates the $\textit{eff}$-$\beta$ curves of various methods for the positive curvature case, while the right plot of Figure \ref{fig:HMC_10} b) shows the negative curvature case. Similar to the previous example, it can be concluded from Table \ref{tab:4a} and \ref{tab:4b} and Figure \ref{fig:HMC_10} that the HMC-SS approaches are as accurate as the conventional CW-MH-SS approach, while the efficiency of RS-HMC-SS is noticeably higher than CW-MH-SS for low probability levels.
%
\begin{table}[!t]
\renewcommand{\thetable}{\arabic{table}.a)}
\centering
\caption{Performance of HMC-SS for nonlinear problem.}
\begin{tabular}{*8c}
\toprule
$\kappa$ &  \multicolumn{3}{c}{RS-HMC} & \multicolumn{3}{c}{CW-MH} & Exact \\ \midrule
  {} & $\hat P_f$   & c.o.v.    & NG   & $\hat P_f$ & c.o.v & NG & $P_f$\\
    0.2 & $6.46\times10^{-5}$ & 0.32 & 4547 & 6.60$\times10^{-5}$ & 0.37 & 4517 & 6.41$\times10^{-5}$\\
    0.6 & $1.40\times10^{-3}$ & 0.26 & 2906 & 1.42$\times10^{-3}$ & 0.28 & 2933 & 1.41$\times10^{-3}$\\
    1.0 & $8.97\times10^{-3}$ & 0.19 & 2566 & 9.02$\times10^{-3}$ & 0.21 & 2550 & 8.99$\times10^{-3}$\\
   -1.0 & $1.37\times10^{-5}$ & 0.38 & 4825 & 1.39$\times10^{-5}$ & 0.42 & 4850 & 1.37$\times10^{-5}$\\
   -5.0 & $6.71\times10^{-6}$ & 0.48 & 5384 & 6.80$\times10^{-6}$ & 0.56 & 5325 & 6.62$\times10^{-6}$\\
 -10.0 & $4.70\times10^{-6}$ & 0.56 & 5441 & 5.20$\times10^{-6}$ & 0.81 & 5433 & 4.73$\times10^{-6}$\\
              \bottomrule
\end{tabular}
\label{tab:4a}
\bigskip
\addtocounter{table}{-1}
\renewcommand{\thetable}{\arabic{table}.b)}
\centering
\caption{Performance of HMC-SS for nonlinear problem.}
\begin{tabular}{*8c}
\toprule
$\kappa$ &  \multicolumn{3}{c}{BB-HMC (Newton)} & \multicolumn{3}{c}{BB-HMC (Secant)} & Exact \\ \midrule
  {} & $\hat P_f$   & c.o.v.    & NG   & $\hat P_f$ & c.o.v & NG & $P_f$\\
    0.2 & $6.47\times10^{-5}$ & 0.23 & 15089 & 6.45$\times10^{-5}$ & 0.22 & 20012 & 6.41$\times10^{-5}$\\
    0.6 & $1.41\times10^{-3}$ & 0.17 & 7132 & 1.42$\times10^{-3}$ & 0.19 & 9232 & 1.41$\times10^{-3}$\\
    1.0 & $9.03\times10^{-3}$ & 0.14 & 6425 & 8.98$\times10^{-3}$ & 0.16 & 8030 & 8.99$\times10^{-3}$\\
   -1.0 & $1.35\times10^{-5}$ & 0.27 & 16066 & 1.36$\times10^{-5}$ & 0.26 & 21314 & 1.37$\times10^{-5}$\\
   -5.0 & $6.60\times10^{-6}$ & 0.29 & 19384 & 6.70$\times10^{-6}$ & 0.33 & 27449 & 6.62$\times10^{-6}$\\
 -10.0 & $4.70\times10^{-6}$ & 0.31 & 19747 & 4.70$\times10^{-6}$ & 0.38 & 28061 & 4.73$\times10^{-6}$\\
              \bottomrule
\end{tabular}
\label{tab:4b}
\end{table}

\break \newpage
\subsubsection{Random vibration example}
\noindent 
Consider a single degree of freedom (SDOF) linear oscillator under seismic loading defined by the differential equation
\begin{equation}
m\ddot{X}(t)+c\dot{X}(t)+kX(t)=-m\ddot{U}_g(t)
\label{eq:rv}
\end{equation}
where $X(t)$, $\dot X(t)$ and $\ddot X(t)$ denote the displacement, velocity and acceleration of the oscillator, respectively. We set the mass $m = 6\times 10^4$[kg], stiffness $k=2.0\times10^7$[N/m], damping $c=2m\zeta\sqrt{k/m}$ with the viscous damping ratio $\zeta = 10\%$. The initial natural period of this SDOF oscillator is $T=0.34$[s]. The ground acceleration $\ddot{U}_g(t)$ is modeled by white noise process. The white noise process is discretized in frequency domain as \cite{Shinozuka:1991aa}
\begin{equation}
\ddot{U}_g(t)=\sigma \sum_{j=1}^{n/2}[u_j\cos(\omega_j t)+\bar u_j\sin(\omega_j t)]
\label{eq:rv_sim}
\end{equation}
in which $u_j$, $\bar u_j$ are independent standard normal random variables, the frequency point is given by $\omega_j = j\Delta\omega$ with a total $n/2=100$ frequency points, the cut-off frequency is set to  $\omega_{cut}=15\pi$, and $\sigma = \sqrt{2S\Delta\omega}$, where $S=0.01[\text{m}^2/\text{s}^3]$ is the intensity of the white noise. The total number of random variables is $n=200$. 

Now we consider the first-passage probability $\Pr[\max_{t\in(0,10)}X(\vect u,t)]>x$. The first passage probabilities for threshold $x=0.020$[m], $x=0.025$[m] and $x=0.030$[m] are computed using Subset Simulation, and the results are compared with the solution obtained from crude MCS with $1.0\times10^6$ runs. Table  \ref{tab:5} illustrates the results. As with the previous examples, Table \ref{tab:5} illustrates that the efficiency of RS-HMC is noticeably higher than CW-MH for low probability levels. 

\begin{table}[!b]
\caption{Performance of HMC-SS for first-passage problem.}
\begin{tabular}{c c c c c c c c c c c}
\toprule
thr.  &  \multicolumn{3}{c}{RS-HMC} & \multicolumn{3}{c}{BB-HMC (Secant)} & \multicolumn{3}{c}{CW-MH} &  MCS\\ \midrule
  {} & $\hat P_f$   & c.o.v.    & NG   & $\hat P_f$ & c.o.v & NG & $\hat P_f$   & c.o.v.    & NG & $P_f$ \\
 \small{0.020} & \small{$6.73\times10^{-3}$} & \small{0.21} & \small{2773} & \small{6.91}$\small{\times10^{-3}}$ & \small{0.15} & \small{10211} & \small{6.67}$\small{\times10^{-3}}$&\small{0.21}&\small{2782}&\footnotesize{6.8}$\footnotesize{\times10^{-3}}$ \\
\small{0.025} & \small{$7.65\times10^{-5}$} & \small{0.32} & \small{4492} & \small{8.91}$\small{\times10^{-2}}$ & \small{0.25} & \small{18551} & \small{7.55}$\small{\times10^{-5}}$&\small{0.35}&\small{4483}&\footnotesize{8.2}$\footnotesize{\times10^{-5}}$ \\
\small{0.030} & \small{$2.90\times10^{-5}$} & \small{0.42} & \small{6400} & \small{3.36}$\small{\times10^{-7}}$ & \small{0.32} & \small{29653} & \small{3.14}$\small{\times10^{-5}}$&\small{0.58}&\small{6400}&$-$ \\
              \bottomrule
\end{tabular}
\label{tab:5}
\end{table}
\subsubsection{Reliability example with elliptical limit-state function}
\noindent
Consider an elliptical limit-state function defined by
\begin{equation}
G(x,y) = r^2 -\frac{(x\cos\theta +y \sin\theta)^2}{c_1^2}-\frac{(x\sin\theta -y \cos\theta)^2}{c_2^2}
\label{eq:lsf_el}
\end{equation}
where $c_1$, $c_2$ and $\theta$ are parameters to control the shape the elliptical function, $r$ directly affects the failure probability, and variables $(x,y)$ follow the bivariate banana shaped distribution introduced in Section \ref{sec:B_HMC} with the same parameter setting. 

\begin{figure}[!h]
    \centering
    \includegraphics[width=0.82\textwidth]{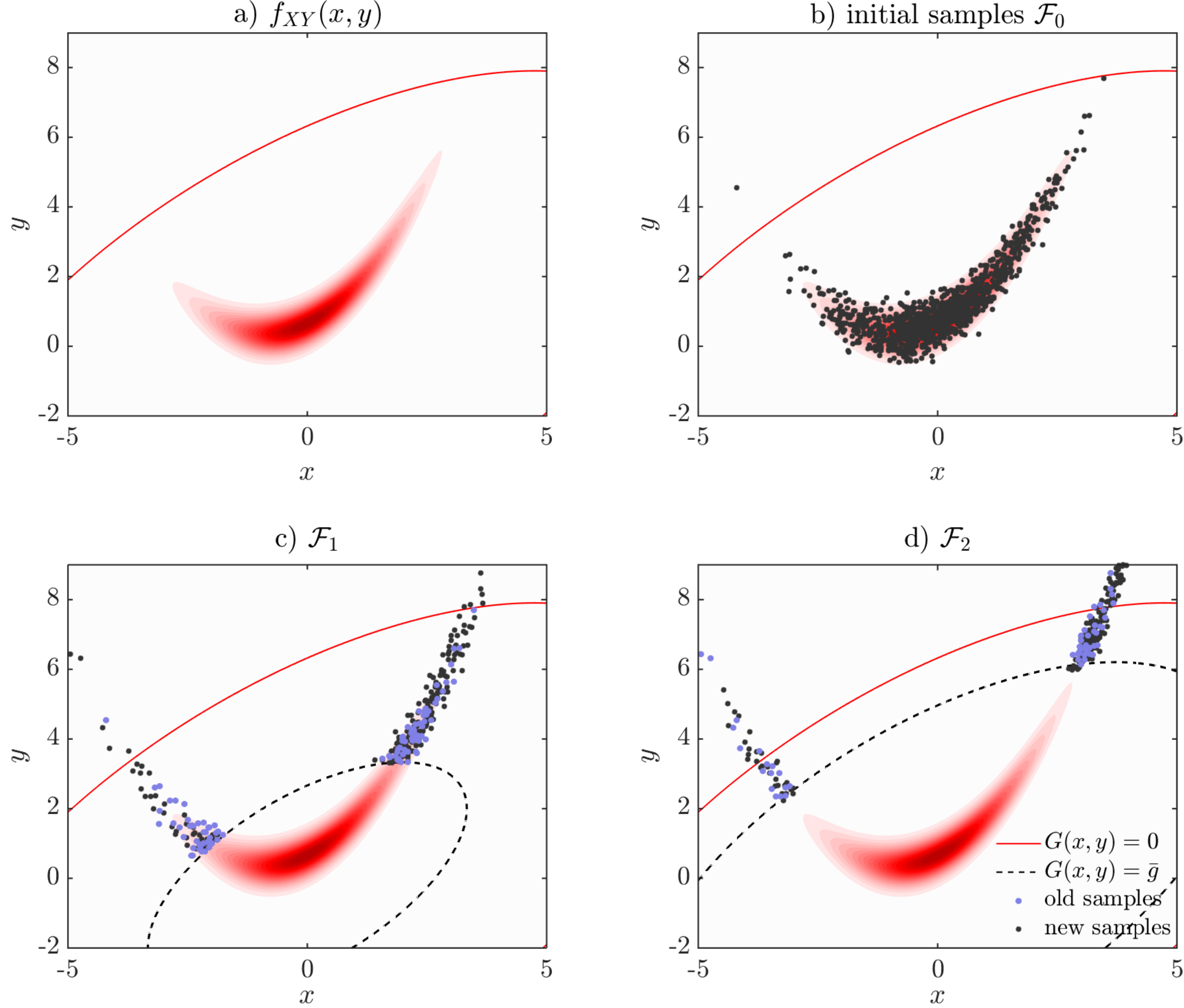}
    \caption{a) The elliptical limit-state function in the space of the banana-shaped distribution; b) initial sampling; c) first subset; d) second subset
    }
    \label{fig:HMC_11}
\end{figure}

\begin{figure}[!h]
    \centering
    \includegraphics[width=0.65\textwidth]{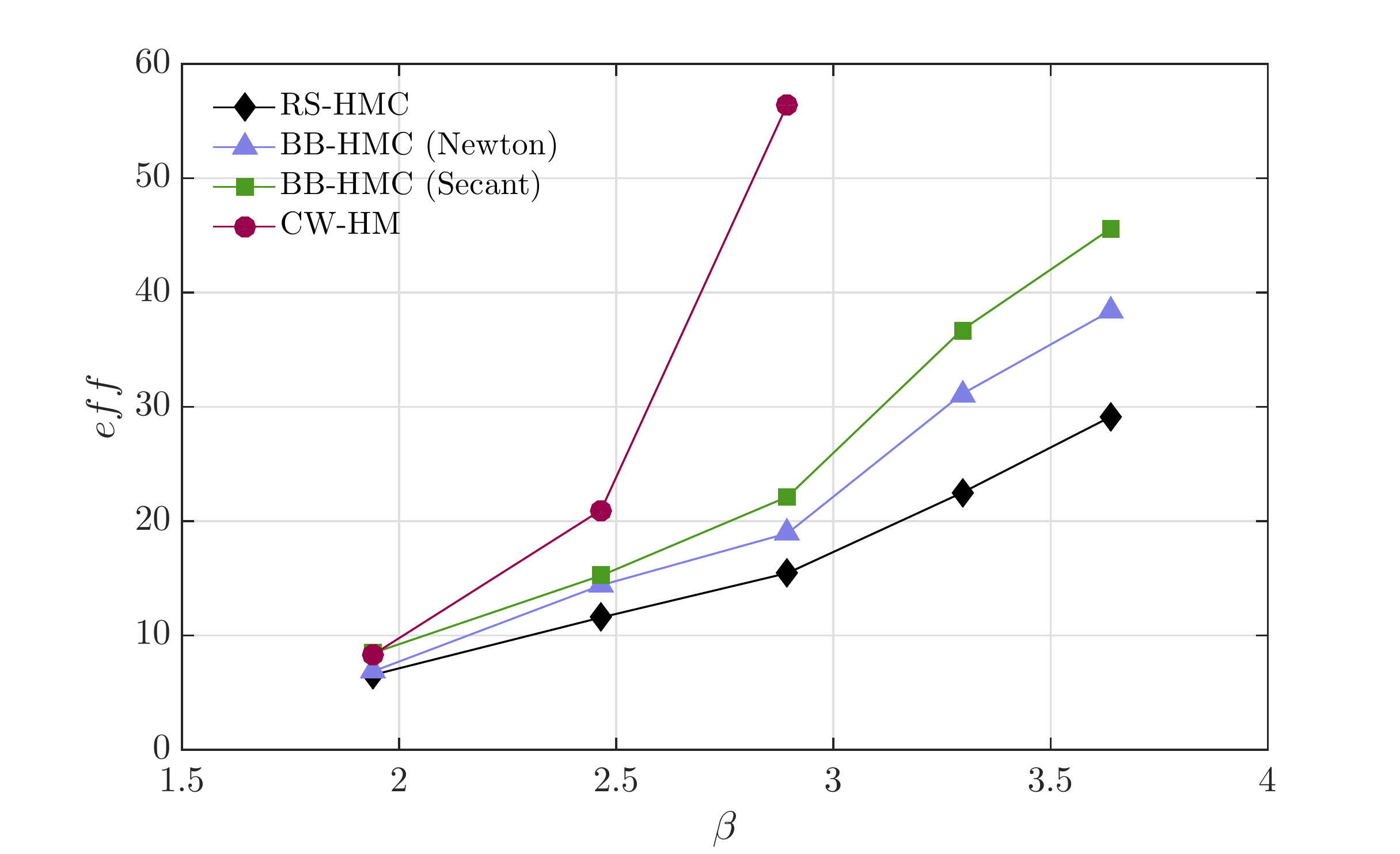}
    \caption{The \textit{eff}-$\beta$ curves for various methods
    }
    \label{fig:HMC_12}
\end{figure}

In this example, parameters in Eq.\eqref{eq:lsf_el} are set as $c_1=1$, $c_2=0.5$ and $\theta=\pi/4$. A uniform transition distribution within a square of width 1 is used in the Metropolis Hastings algorithm. For the leapfrog based RS-HMC, $\Delta t= 0.05$, $L=\text{round}(t_f/\Delta t)$ and $t_f$ is selected adaptively using Algorithm \ref{alg:mean_T}. Table \ref{tab:6a} and Table \ref{tab:6b} illustrate the results obtained from RS-HMC-SS, BB-HMC-SS and MH-SS approaches compared with the solution obtained from crude MCS with $1.0\times 10^{6}$ runs for a sequence of $r$ values. Note that we do not obtain meaningful results of MH-SS for $r>10$. This suggests that setting a constant transition distribution in MH algorithm is not suitable for this example. On the other hand, the performance of HMC-SS approaches is as robust as that in previous examples formulated in Gaussian space. From Figure \ref{fig:HMC_11}, it is clear that contributions to the probability of failure arise from both tails of the joint distribution. It is interesting to note that the HMC-SS is able to explore both tails efficiently. This is due to the fact that the narrow probability spaces are better explored by Hamiltonian dynamics compared to simple MCMC. Figure \ref{fig:HMC_12} illustrates the \textit{eff}-$\beta$ curves of various methods.
\begin{table}[!b]
\renewcommand{\thetable}{\arabic{table}.a)}
\centering
\caption{Performance of HMC-SS for elliptical limit-state problem.}
\begin{tabular}{*8c}
\toprule
$r$ &  \multicolumn{3}{c}{RS-HMC} & \multicolumn{3}{c}{CW-MH} & Exact \\ \midrule
  {} & $\hat P_f$   & c.o.v.    & NG   & $\hat P_f$ & c.o.v & NG & $P_f$\\
    6 & $2.64\times10^{-2}$ & 0.15 & 1900 & 2.52$\times10^{-5}$ & 0.19 & 1900 & 2.63$\times10^{-2}$\\
    8 & $6.88\times10^{-3}$ & 0.22 & 2769 & 6.20$\times10^{-5}$ & 0.40 & 2750 & 6.85$\times10^{-3}$\\
  10 & $1.82\times10^{-3}$ & 0.29 & 2843 & 1.61$\times10^{-5}$ & 0.75 & 5675 & 1.90$\times10^{-3}$\\
  12 & $4.87\times10^{-4}$ & 0.37 & 3691 &  $-$ 			  & $-$   &  $-$   & 4.91$\times10^{-4}$\\
  14 & $1.36\times10^{-4}$ & 0.46 & 4024 & $-$ 			  & $-$   &  $-$   & 1.36$\times10^{-4}$\\
              \bottomrule
\end{tabular}
\label{tab:6a}
\bigskip
\addtocounter{table}{-1}
\renewcommand{\thetable}{\arabic{table}.b)}
\centering
\caption{Performance of HMC-SS for elliptical limit-state problem.}
\begin{tabular}{*8c}
\toprule
$r$ &  \multicolumn{3}{c}{BB-HMC (Newton)} & \multicolumn{3}{c}{BB-HMC (Secant)} & Exact \\ \midrule
  {} & $\hat P_f$   & c.o.v.    & NG   & $\hat P_f$ & c.o.v & NG & $P_f$\\
    6 & $2.63\times10^{-2}$ & 0.12 & 3244& 2.66$\times10^{-2}$ & 0.13 & 4218 & 2.63$\times10^{-2}$\\
    8 & $6.90\times10^{-3}$ & 0.20 & 5185 & 7.00$\times10^{-3}$ & 0.21 & 5269 & 6.85$\times10^{-3}$\\
  10 & $2.00\times10^{-3}$ & 0.26 & 5309 & 1.80$\times10^{-3}$ & 0.26 & 7250 & 1.90$\times10^{-3}$\\
  12 & $4.75\times10^{-4}$ & 0.35 & 7903 & 4.60$\times10^{-4}$ & 0.36 &  10406  & 4.91$\times10^{-4}$\\
  14 & $1.25\times10^{-4}$ & 0.41 & 8793 & 1.39$\times10^{-4}$ & 0.43 & 11271 & 1.36$\times10^{-4}$\\
              \bottomrule
\end{tabular}
\label{tab:6b}
\end{table}
%
%

Finally, it is of interest to note that the aforementioned results of this example are obtained using independent and identically distributed (i.i.d.) samples as seeds for the initial subset. However, for complex distribution models in which one cannot directly generate i.i.d. samples, one may use MCMC algorithm to generate samples for the initial subset. For such cases, the c.o.v. of the failure probability estimate may increase due to the inherent correlation of the initial subset samples. It turns out, compared to traditional MH algorithms, HMC is particularly suitable to resolve the correlation issue since HMC typically has a short burn-in phase and a low autocorrelation lag. 

\begin{figure}[!b]
    \centering
    \includegraphics[width=0.85\textwidth]{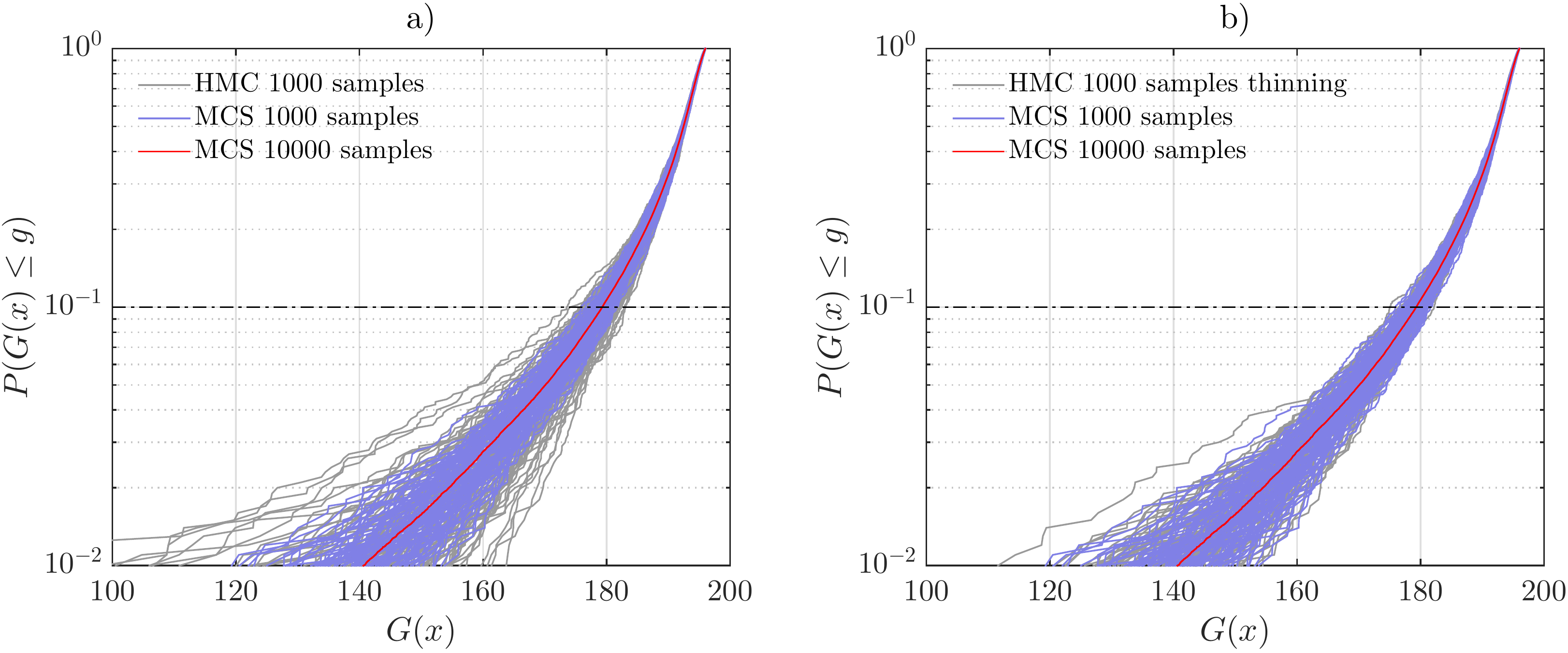}
    \caption{a) Hundred CDFs obtained from 1000 samples of un-thinned HMC chain vs hundred CDFs from 1000 MCSs samples. b) Hundred CDFs obtained from 1000 samples of a thinned HMC chain of length 10,000 and thinning lag 10 vs hundred CDFs from 1000 MCSs samples. Dashed line represent the 0.1 percentile.
    }
    \label{fig:HMC_13}
\end{figure}

To reduce the inherent correlation of the initial subset samples, one could use HMC combined with a thinning procedure  \cite{link2012thinning, maceachern1994subsampling}. Specifically, it is recommended to thin the chain by subsampling every $k$ samples, where $k$ denotes the thinning lag. In principle, the thinning lag should be selected larger than the autocorrelation lag of the chain, so that the correlation between the samples is effectively reduced. Figure \ref{fig:HMC_13} shows the effect of thinning in the initial adaptive limit-state surface of this example. The thinning lag is 10, and it is chosen based on a study of the autocorrelation lag of an HMC chain. Figure \ref{fig:HMC_13} a) shows, in grey, the CDFs (defined as $\Pr[G(x,y)\le g]$  of 100 un-thinned chains of length 1000, compared to 100 CDFs obtained from a classical MCS with uncorrelated samples (reported in light blue). As a reference, in red color, it is reported the CDF obtained with a MCS based on 10,000 samples. Figure \ref{fig:HMC_13} b) shows, in gray, the CDFs of 100 thinned chains of length 1000 sampled from chains of length 10,000. The MCSs in light blue and red are the same of Figure \ref{fig:HMC_13}a). From this Figure, it is evident that the samples obtained by thinning are almost equivalent to the uncorrelated ones obtained by MCS. Table \cite{Au:2001aa} reports the failure probability estimate and the c.o.v. for $r=14$, for a sequence of thinning lengths, obtained from RS-HMC-SS. It is shown in Table \ref{tab:7a} and \ref{tab:7b} that the c.o.v, generally, decreases with the increase of the thinning lag. For this example, a thinning lag of $k=3$ noticeably reduces the c.o.v. For a lag $k\ge5$, a significant decrease in c.o.v is not observed since the samples are already almost uncorrelated. Notice that the estimated c.o.v. of lag 10 is higher than the estimated c.o.v of lag 5 only because of the inherent variability of the statistical estimates. Note that thinning is applied to the initial subset only, and only after the thinning the limit-state function is evaluated for each thinned sample, thus the thinning procedure does not introduce additional limit-state function evaluations.

The reader should be aware that the thinning is not, in general, a recommended practice for approximating means, variances or percentiles. It is often better to use the full correlated chain rather than the thinned de-correlated one \cite{link2012thinning, maceachern1994subsampling}. However, in the context of Subset Simulation, the number of samples used to evaluate the conditional probability per subset is fixed. Therefore, in this particular context, for the fist subset, using a thinned chain of $N$ is expected to be more effective than using an un-thinned chain of $N$ samples. 
%

\begin{table}[!h]
\renewcommand{\thetable}{\arabic{table}.a)}
\centering
\caption{Thinning effect on the c.o.v of the failure probability estimate.}
\begin{tabular}{c c c c c c c }
\toprule 
$r$  &  \multicolumn{2}{c}{RS-HMC} & \multicolumn{2}{c}{RS-HMC} & \multicolumn{2}{c}{RS-HMC}\\ 
{ } & \multicolumn{2}{c}{i.i.d.} & \multicolumn{2}{c}{$(k=0)$ } & \multicolumn{2}{c}{$(k=3)$ }  \\ 
\midrule
  {} & $\hat P_f$   & c.o.v.    & $\hat P_f$   & c.o.v.  & $\hat P_f$  & c.o.v.  \\
  {14} &  {$1.36\times10^{-3}$} &  {0.46} &   {$1.36\times10^{-3}$} &  {0.71} &   {$1.36\times10^{-3}$} &0.54 \\
              \bottomrule
\end{tabular}
\label{tab:7a}
\bigskip
\addtocounter{table}{-1}
\renewcommand{\thetable}{\arabic{table}.b)}
\centering
\caption{Thinning effect on the c.o.v of the failure probability estimate.}
\begin{tabular}{c c c c c c c }
\toprule
$r$  &  \multicolumn{2}{c}{RS-HMC} & \multicolumn{2}{c}{RS-HMC} & \multicolumn{2}{c}{RS-HMC}\\ 
{ } & \multicolumn{2}{c}{i.i.d.} &   \multicolumn{2}{c}{ $(k=5)$} & \multicolumn{2}{c}{$(k=10)$}  \\ 
\midrule
  {} & $\hat P_f$   & c.o.v.    & $\hat P_f$   & c.o.v.  & $\hat P_f$  & c.o.v.  \\
  {14} &  {$1.36\times10^{-3}$} &  {0.46} &  {1.29}$ {\times10^{-3}}$& {0.44}& {1.36}$ {\times10^{-3}}$&0.47 \\
              \bottomrule
\end{tabular}
\label{tab:7b}
\end{table}
\subsubsection{Reliability example of pushover analysis of a shear-frame structure}
\noindent
Consider a push over analysis of the three stories frame in Figure 14. The interstory behavior is inelastic with a force-interstory-drift relationship based on a $J2$ plasticity model \cite{simo2006computational}. Both kinematic and isotropic hardenings are considered nulls; therefore, the model can be regarded as elastic-perfectly plastic; then, the elastic domain is completely defined by the parameter $u_y$, which is the yielding displacement. It is assumed the horizontal forces are deterministic and known. The initial inter-story stiffnesses are considered correlated lognormal random variables. The horizontal force values, the means of the stiffnesses, c.o.vs, and correlation coefficients are reported in Table \ref{tab:8}.  The limit state function is defined as 
$g(x,\vect v)=\max(\vect v)$, where $\vect v=[v_1,v_2,v_3]$, with  $v_1, v_2$ and $v_3$ being the first, second and third interstory drift. The threshold $x=0.12$[m] corresponds to $3\%$ of the interstory height, which is assumed to be $4$[m]. In this example we focus solely on leapfrog based RS-HMC-SS directly applied in the original probability space. The BB-HMC-SS is not considered because the derivative of the limit state function either does not exist (there is a discontinuity between the elastic and the plastic domains), or it is difficult to obtain. The results are compared with MH-SS using a uniform transition distribution within a square of width 1, and crude MCS based on $10^5$ runs. The results, reported in Table  \ref{tab:9}, show that RS-HMC-SS in the original space provides both an accurate and efficient reliability estimate of the system. Similar to the previous examples, Table \ref{tab:9} confirms that the efficiency of RS-HMC-SS is noticeably higher than MH-SS.
\begin{figure}[h]
    \centering
    \includegraphics[width=0.6\textwidth]{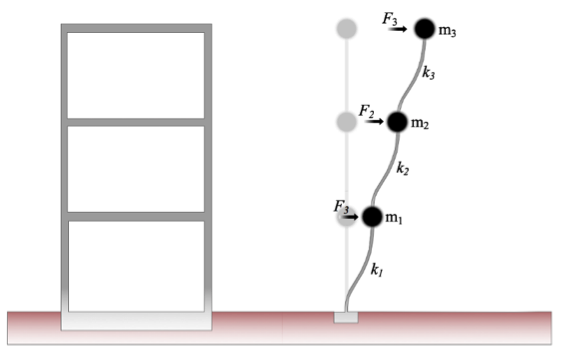}
    \caption{Structural archetype
    }
    \label{fig:HMC_14}
\end{figure}
\begin{table}[!h]
\centering
\caption{Structural and load properties; $\mu$ represents the mean value.}
\begin{tabular}{*6c}
\toprule
{} & $\mu_k$[N/m] & c.o.v. & $\rho$ &$u_y$ &$F$[N] \\ \midrule
Story 1 & $3.0\times10^{8}$ & 0.1 & $\rho_{1-2}=0.6$ & $0.04$&$1.645\times10^{8}$\\
Story 2 & $2.8\times10^{8}$ & 0.1 & $\rho_{2-3}=0.6$ & $0.04$&$2.585\times10^{8}$\\
Story 3 & $1.5\times10^{8}$ & 0.1 & $\rho_{3-1}=0.6$ & $0.04$&$4.700\times10^{8}$\\
              \bottomrule
\end{tabular}
\label{tab:8}
\end{table}
\begin{table}[!h]
\centering
\caption{Performance of HMC-SS for structural reliability of a push over analysis.}
\begin{tabular}{*9c}
\toprule
$x$ &  \multicolumn{3}{c}{RS-HMC} & \multicolumn{3}{c}{CW-MH}  & \multicolumn{2}{c}{MCS}  \\ \midrule
 {} & $\hat P_f$   & c.o.v.    & NG   & $\hat P_f$ & c.o.v & NG & $P_f$& c.o.v.\\
0.12 & $2.78\times10^{-4}$ & 0.33 & 3700 & $2.81\times10^{-4}$ &0.41& 3700 &$2.60\times10^{-4}$& 0.20\\

              \bottomrule
\end{tabular}
\label{tab:9}
\end{table}
%
%
\section{Conclusion}\label{sec:conc}
\noindent
A Hamiltonian Monte Carlo (HMC) approach is developed for Subset Simulation in reliability analysis. The HMC method operates via simulating a deterministic Hamiltonian system to propose samples for a target probability distribution. The method is designed to alleviate the random walk behavior so that a more effective exploration of the probability space can be expected compared to standard Gibbs or Metropolis-Hastings techniques. Two HMC approaches for sampling the inter-mediate conditional probability distributions of Subset Simulation are developed. The first approach relies on rejection sampling, and the other one simulates a bouncing mechanics of the Hamiltonian system when the proposed trajectory interacts with the barrier/constraint of the probability distribution. 

	To study the effectiveness of the proposed HMC approaches, first, the behavior of HMC method is illustrated and tested by simple Gaussian and non-Gaussian probability distribution models. The results confirm that, compared to the traditional random walk Metropolis Hastings approach, a more effective exploration of the probability space can be generally expected from the HMC approaches. Next, the performance of the two HMC based Subset Simulation methods is tested using reliability examples with explicit linear and nonlinear limit-state functions, a random vibration example, and two reliability problems formulated in non-Gaussian spaces. The numerical results indicate that the two HMC approaches are at least as accurate as the conventional Metropolis Hastings approach, while the efficiency of the rejection sampling based HMC approach is noticeably higher than the conventional approach. For the barrier bouncing based HMC (BB-HMC) approach, due to the additional computational cost required in determining the time point (hitting time) at which the Hamiltonian system interacts with the barrier/constraint using a secant or Newton-Raphson algorithm, it seems that the application of the method in Subset Simulation for general reliability problems is not attractive. However, for problems with simple explicit limit-state functions, e.g., response surface based reliability analysis where analytical surrogate limit-state function is available, one may find analytical solution for the hitting time in BB-HMC and consequently the additional computational cost in BB-HMC can be eliminated. In this case, the BB-HMC based Subset Simulation outperforms all the other Subset Simulation approaches studied in this paper.  

	It is concluded from this study that HMC method provides an attractive and also general alternative to perform MCMC sampling in Subset Simulation. Finally, for reliability problems formulated in high dimensional and strong correlated probability space, to further improve the efficiency of HMC based Subset Simulation, it may of interest to investigate the use of a variant HMC method which incorporates the geometrical information of the probability space, namely Riemann manifold Hamiltonian Monte Carlo method. Another promising line of application for the current framework is the Bayesian analysis of reliability under parameter uncertainties. In this setting, the reliability of the system depends upon a set of random parameters which value is estimated and updated through Bayesian inference. HMC offers the perfect setting to sample from complex posterior distributions of the parameter distributions enabling, therefore, an efficient up-date of the reliability of the mechanical system.  

\section*{Acknowledgement}
\noindent
Dr. Ziqi Wang was supported by the National Science and Technology Major Project of the Ministry of Science and Technology of China (Grant No. 2016YFB0200605). Dr. Marco Broccardo was supported by the Swiss Competence Center for Energy Research Supply of Electricity. Professor  Junho Song was supported by the Institute of Construction and Environmental Engineering at Seoul National University, and the project ``Development of Lifecycle Engineering Technique and Construction Method for Global Competitiveness Upgrade of Cable Bridges'' funded by the Ministry of Land, Infrastructure and Transport (MOLIT) of the Korean Government (Grant No. 16SCIP-B119960-01). This support is gratefully acknowledged. Any opinions, findings, and conclusions expressed in this paper are those of the authors, and do not necessarily reflect the views of the sponsors.


\bibliographystyle{elsarticle-num} 
\bibliography{HMC.bib}

\begin{thebibliography}{10}
\expandafter\ifx\csname url\endcsname\relax
  \def\url#1{\texttt{#1}}\fi
\expandafter\ifx\csname urlprefix\endcsname\relax\def\urlprefix{URL }\fi
\expandafter\ifx\csname href\endcsname\relax
  \def\href#1#2{#2} \def\path#1{#1}\fi

\bibitem{ditlevsen1996str}
O.~Ditlevsen, H.~O. Madsen, Structural reliability methods, Vol. 178, Wiley New
  York, 1996.

\bibitem{der2005first}
A.~Der~Kiureghian, et~al., First-and second-order reliability methods,
  Engineering design reliability handbook (2005) 14--11.

\bibitem{faravelli1989resp}
L.~Faravelli, Response-surface approach for reliability analysis, Journal of
  Engineering Mechanics 115~(12) (1989) 2763--2781.

\bibitem{bucher1990fast}
C.~G. Bucher, U.~Bourgund, A fast and efficient response surface approach for
  structural reliability problems, Structural Safety 7~(1) (1990) 57--66.

\bibitem{rubinstein1998modern}
R.~Y. Rubinstein, B.~Melamed, Modern simulation and modeling, Vol.~7, Wiley New
  York, 1998.

\bibitem{rubinstein2013cross}
R.~Y. Rubinstein, D.~P. Kroese, The cross-entropy method: a unified approach to
  combinatorial optimization, {M}onte-{C}arlo {S}imulation and machine
  learning, Springer Science \&amp; Business Media, 2013.

\bibitem{Au:2001aa}
S.~K. Au, J.~L. Beck, Estimation of small failure probabilities in high
  dimensions by subset simulation, Probababilistic Engineering Mechanics 16~(4)
  (2001) 263--277.

\bibitem{kurtz2013cross}
N.~Kurtz, J.~Song, Cross-entropy-based adaptive importance sampling using
  {G}aussian mixture, Structural Safety 42 (2013) 35--44.

\bibitem{cerou2012sequential}
F.~C{\'e}rou, P.~Del~Moral, T.~Furon, A.~Guyader, Sequential {M}onte {C}arlo
  for rare event estimation, Statistics and Computing 22~(3) (2012) 795--808.

\bibitem{neal2001annealed}
R.~M. Neal, Annealed importance sampling, Statistics and computing 11~(2)
  (2001) 125--139.

\bibitem{miao2011modified}
F.~Miao, M.~Ghosn, Modified subset simulation method for reliability analysis
  of structural systems, Structural Safety 33~(4) (2011) 251--260.

\bibitem{papaioannou2014mcmc}
I.~Papaioannou, W.~Betz, K.~Zwirglmaier, D.~Straub, {MCMC} algorithms for
  subset simulation, Probabilistic Engineering Mechanics 41 (2015) 89--103.

\bibitem{duane1987hybrid}
S.~Duane, A.~D. Kennedy, B.~J. Pendleton, D.~Roweth, Hybrid {M}onte {C}arlo,
  Physics letters B 195~(2) (1987) 216--222.

\bibitem{neal2011mcmc}
R.~M. Neal, et~al., {MCMC} using hamiltonian dynamics, Handbook of Markov Chain
  Monte Carlo 2 (2011) 113--162.

\bibitem{neal1993probabilistic}
R.~M. Neal, Probabilistic inference using {M}arkov chain {M}onte {C}arlo
  methods, Tech. rep., Department of Computer Science, University of Toronto
  Toronto, Ontario, Canada (1993).

\bibitem{neal2012bayesian}
R.~M. Neal, Bayesian learning for neural networks, Vol. 118, Springer Science
  \& Business Media, 2012.

\bibitem{akhmatskaya2008gshmc}
E.~Akhmatskaya, S.~Reich, {GSHMC}: {A}n efficient method for molecular
  simulation, Journal of Computational Physics 227~(10) (2008) 4934--4954.

\bibitem{strathmann2015gradient}
H.~Strathmann, D.~Sejdinovic, S.~Livingstone, Z.~Szabo, A.~Gretton,
  Gradient-free {H}amiltonian {M}onte {C}arlo with efficient kernel exponential
  families, in: Advances in Neural Information Processing Systems, 2015, pp.
  955--963.

\bibitem{au2016mcmc}
S.-K. Au, On {MCMC} algorithm for subset simulation, Probabilistic Engineering
  Mechanics 43 (2016) 117--120.

\bibitem{haario2001adaptive}
H.~Haario, E.~Saksman, J.~Tamminen, An adaptive {M}etropolis algorithm,
  Bernoulli (2001) 223--242.

\bibitem{horowitz1991generalized}
A.~M. Horowitz, A generalized guided {M}onte {C}arlo algorithm, Physics Letters
  B 268~(2) (1991) 247--252.

\bibitem{pakman2014exact}
A.~Pakman, L.~Paninski, Exact {H}amiltonian {M}onte {C}arlo for truncated
  multivariate gaussians, Journal of Computational and Graphical Statistics
  23~(2) (2014) 518--542.

\bibitem{zhang1993dynamic}
Y.~Zhang, A.~Der~Kiureghian, Dynamic response sensitivity of inelastic
  structures, Computer Methods in Applied Mechanics and Engineering 108~(1-2)
  (1993) 23--36.

\bibitem{verlet1967computer}
L.~Verlet, Computer ``experiments" on classical fluids. i. {T}hermodynamical
  properties of {L}ennard-{J}ones molecules, Physical review 159~(1) (1967) 98.

\bibitem{hoffman2014no}
M.~D. Hoffman, A.~Gelman, The no-{U}-turn sampler: adaptively setting path
  lengths in {H}amiltonian {M}onte {C}arlo., Journal of Machine Learning
  Research 15~(1) (2014) 1593--1623.

\bibitem{au2012discussion}
S.~Au, J.~L. Beck, K.~M. Zuev, L.~S. Katafygiotis, Discussion of paper by {F}.
  {M}iao and {M}. {G}hosn ``{M}odified subset simulation method for reliability
  analysis of structural systems'', {S}tructural {S}afety, 33: 251--260, 2011,
  Structural Safety 34~(1) (2012) 379--380.

\bibitem{au2007application}
S.~Au, J.~Ching, J.~Beck, Application of subset simulation methods to
  reliability benchmark problems, Structural Safety 29~(3) (2007) 183--193.

\bibitem{Shinozuka:1991aa}
D.~G. Shinozuka~M, Simulation of stochastic processes by spectral
  representation., ASME. Appl. Mech. Rev. 44~(4) (1991) 191--204.

\bibitem{link2012thinning}
W.~A. Link, M.~J. Eaton, On thinning of chains in {MCMC}, Methods in Ecology
  and Evolution 3~(1) (2012) 112--115.

\bibitem{maceachern1994subsampling}
S.~N. MacEachern, L.~M. Berliner, Subsampling the {G}ibbs sampler, The American
  Statistician 48~(3) (1994) 188--190.

\bibitem{simo2006computational}
J.~C. Simo, T.~J. Hughes, Computational inelasticity, Vol.~7, Springer Science
  \& Business Media, 2006.

\end{thebibliography}


%
%
%
%
\end{document}